\documentclass[lineno]{jfm}

\usepackage[outdir=./]{epstopdf}
\usepackage{graphicx}
\usepackage{newtxtext}
\usepackage{newtxmath}
\usepackage{natbib}
\usepackage{hyperref}
\usepackage{bbm, amsmath}
\usepackage{tikz,pgfplots, float}
\usepackage{tikz-3dplot}

\hypersetup{
    colorlinks = trueJFM,
    urlcolor   = blue,
    citecolor  = black,
}

\newcommand{\grad}{\boldsymbol{\nabla}} 
\newcommand{\wave}{\mathbbm{B}}
\newcommand{\pair}{\mathbbm{T}}
\newcommand{\domain}{\tilde{\mathbbm{B}}}
\newcommand{\Mz}{\mathcal{M}_{0\textrm{D}}}
\newcommand{\Mt}{\mathcal{M}_{2\textrm{D}}}

\newcommand{\RomanNumeralCaps}[1]
\linenumbers

\shorttitle{The long view of TRI in finite-width internal wave beams}
\shortauthor{K. M. Grayson, Stuart B. Dalziel and Andrew G. W. Lawrie}

\title{The long view of triadic resonance instability in finite-width internal gravity wave beams} 

\author{K. M. Grayson\aff{1}
	\corresp{\email{katherinegrayson@icloud.com}},
	Stuart B. Dalziel\aff{1}
	\and Andrew G. W. Lawrie\aff{2}}

\affiliation{\aff{1}Department of Applied Mathematics and Theoretical Physics, University of Cambridge,
	Wilberforce Road, Cambridge, CB3 0WA, UK 
	\aff{2}Department of  Engineering, University of
	Bristol, Academic Street, Camford CF3 5QL, UK}

\begin{document}
\maketitle

\begin{abstract}
	
	
	This paper presents our investigation into the modification of a finite-width internal gravity wave beam arising from triadic resonance instability. We present both experimental and weakly non-linear modelling to examine this instability mechanism, in which a primary wave beam generates two secondary wave beams of lower frequencies and shorter length scales. Through a versatile experimental set-up, we examine how this instability develops over hundreds of buoyancy periods. Unlike predictions from previous zero-dimensional weakly non-linear theory, we find that the approach to a saturated equilibrium state for the triadic interactions is not monotonic; rather, the amplitudes and structures of the constituent beams continue to modulate without ever reaching a steady equilibrium. To understand this behaviour we develop a weakly non-linear approach to account for the spatio-temporal evolution of the amplitudes and structures of the beams over slow time-scales and long distances, and explore the consequences using a numerical scheme. Through this approach, we establish that the evolution of the instability is remarkably sensitive to the spatio-temporal triadic configuration for the system and how part of the observed  modulations can be attributed to a competition between the linear growth rate of the secondary wave beams and the finite residence time of the triadic perturbations within the underlying primary beam.
\end{abstract}

\begin{keywords}
	Authors should not enter keywords on the manuscript, as these must be chosen by the author during the online submission process and will then be added during the typesetting process (see \href{https://www.cambridge.org/core/journals/journal-of-fluid-mechanics/information/list-of-keywords}{Keyword PDF} for the full list).  Other classifications will be added at the same time.
\end{keywords}


\section{Introduction}
\label{sec:intro}

The meridional overturning circulation is critical in the regulation of the earth's climate, and understanding the processes essential for maintaining this circulation is of key importance in global climate models. \citet{Munk1966} was amongst the first to suggest that internal gravity waves play a significant part in the deep water vertical mixing of the density stratification within the open ocean, and hence the maintenance of these currents. It is now well established that the breaking of internal waves contributes to the turbulent mixing in the ocean \citep{Staquet2002, Wunsch2004}, yet only recently have the pathways by which internal waves transfer energy to smaller scales and the eventual breaking events been examined in more detail. As noted by \citet{Dauxois2018}, our understanding of these dissipative mechanisms, as opposed to internal wave generation, leaves several open questions. 

Various key mechanisms have been cited for how large-scale internal waves cascade energy to smaller scales. These include internal wave reflection off sloping boundaries \citep{Nash2004}, critical angle reflection \citep{Dauxois1999} and scattering due to small scale topography \citep{Peacock2009}. A review by \cite{Sarkar2017} suggests that no single mechanism is responsible for the internal wave contribution to the energy cascade, rather it is a combination of multiple linear and eventually non-linear processes. \citet{MacKinnon2005} and \citet{Alford2007} suggested that, equatorward of a critical latitude \citep{Richet2018}, parametric sub-harmonic instability (PSI), plays a dominant role in the energy transformation of the internal tide into higher-mode near-inertial waves. Indeed, \citet{Sutherland2013} argues that away from sea-floor boundaries, and neglecting the distorting influence of ocean currents, PSI is one of the primary mechanisms for the energy cascade in the abyssal ocean. PSI can be viewed a special case of triadic resonance instability (TRI), which is a weakly non-linear, slowly-growing resonant mechanism whereby a primary wave becomes unstable due to infinitesimal perturbations within the flow. As the instability grows, a resonant triad interaction forms whereby the primary wave transfers energy to two secondary waves of lower frequency and shorter length scale \citep{Staquet2002}. 

In the inviscid limit and under the assumption of an infinite plane-wave, the frequencies of the secondary waves in the triad are equal to half of the primary wave, motivating the traditional terminology of PSI \citep{Fan2019}. While one often makes the appropriate assumption of oceanic scales being inviscid, in the laboratory setting (where scales are smaller), viscous effects cannot be neglected and resonant wave frequencies deviate away from this sub-harmonic relationship. Moreover, for certain beam widths, the finite-amplitude manifestation of this instability is unable to access these sub-harmonic frequencies \citep{Bourget2014}. In the context of a viscous finite-width beam it is therefore more appropriate to refer to TRI as opposed to PSI.  

The first reported experimental evidence of TRI for internal and interfacial waves was approximately 50 years ago by \citet{Davis1967}, \citet{McEwan1971} and \cite{McEwan1977}, who showed that for finite-width beams there exists an amplitude threshold that must be surpassed for instability to occur. This threshold is not found in the limiting case of an infinite plane-wave, where infinitesimal perturbations may induce the development of the instability \citep{Koudella2006}. In fact, in the special case of a linearly stratified Boussinesq fluid, a plane-waveform holds the peculiar property of being an exact solution to the full non-linear equations at any amplitude \citep[e.g.][]{Thorpe1968, Thorpe1987, Sutherland2006}, albeit not a linearly stable one. However, while single monochromatic plane-waves are convenient mathematically, in nature waves will never take this form. Realistically, oceanic waves are generated from baroclinic tides across ocean ridges and will manifest as beams confined locally in space and therefore broadly distributed over the wavenumber spectrum \citep{Lamb2004, Gostiaux2007}. The focus of analyses using plane-wave solutions has been highlighted in the review by \citet{Dauxois2018}, who argue (correctly in our view) that the effects of finite-width and envelope shape play an important, but generally overlooked role, when considering the non-linearities of internal waves. 


In attempting to address these concerns, researchers have turned towards exploring the dynamics of TRI in spatially localised internal wave beams. Building on the work of \cite{Bourget2013}, \cite{Bourget2014} calculate a growth rate for the instability based on a energy balance that accounts for the role of a finite-width beam. Using direct numerical simulations they also show that the amplitude threshold for instability decreases as the beam width is increased. This decrease is due to any perturbations having a larger spatial field (and hence a longer time) in which to interact with the underlying primary beam, thereby increasing the spatial region (and time interval) over which energy can be transferred. These findings align with the theoretical work of \cite{Karimi2014}, who show how the form of the carrier envelope for a finite-width wave beam has a significant influence on its ability to become unstable based on the wavenumber spectrum produced from the windowing. These works highlight the duality of interpretation for finite-width beams in terms of both the physical parameters and the spectrum in Fourier space.

Triadic resonance can arise due to the sustained spatio-temporal interactions that occur when
\begin{equation}
\label{eqn: resonant_condition}
{\phi_0} = {\phi_1} + {\phi_2}, 
\end{equation}
where the wave phase, $\phi_p$, is defined as 
\begin{equation}
\label{eqn:  phase resonant_condition}
\phi_p = \boldsymbol{k}_p \cdot \boldsymbol{x} - \omega_p t.
\end{equation}
The subscript $p = (0, 1, 2)$ is used throughout this paper to define the primary wave and the two secondary waves, respectively. Both (\ref{eqn: resonant_condition}) and (\ref{eqn:  phase resonant_condition}) are true for three-dimensions, but, without loss of generality we can rotate to a two-dimensional (2D) co-ordinate system. The 2D wave vector of wave $p$ is defined as $\boldsymbol{k}_p = (l_p,m_p)$ with magnitude $|\boldsymbol{k}_p| = \kappa_p$, where the components are given in Cartesian co-ordinates $(x,z)$, marked in Figure \ref{fig: experimental setup}, and $\omega_p$ denotes the frequency of the wave. In order to distinguish between the three wave beams in the triad and their corresponding parameters, we define 
\begin{equation}
\label{eqn: define W}
\wave_p =  \{ \rho_p, \Psi_p; \omega_p, \boldsymbol{k}_p, \Lambda_p ...\},
\end{equation}
where $\wave_p$ indicates a wave beam with density $\rho_p$ and stream function $\Psi_p$ fields, frequency $\omega_p$, characteristic wavenumber vector $\boldsymbol{k}_p$ and beam width $\Lambda_p$. For the primary beam $\omega_0$, $\boldsymbol{k}_0$, and $\Lambda_0$ are imposed control parameters, whereas for the secondary beams they arise from the triadic conditions and (weakly) non-linear dynamics. All triadic wave beams must also satisfy the dispersion relationship for internal waves given as 
%
\begin{equation}
\label{eqn: dispersion}
\frac{\omega_p}{N} = \pm \cos{\theta_p} = \pm\frac{\left|l_p\right|}{\sqrt{{l_p}^2 + {m_p}^2}},
\end{equation}
%
where $\theta_p$ is the angle between the lines of constant phase and the vertical and $l_p$ and $m_p$ are the characteristic wavenumber contributions from each beam. Here $N$ is the buoyancy frequency of the stratification given by 
\begin{equation}
\label{eqn: bouynacy freq}
N = \sqrt{-\frac{g}{ \varrho_0}\frac{\partial \bar{\rho}}{\partial z}},
\end{equation}
where $g$ is the gravitational constant of acceleration. Under the assumptions of a Boussinesq, incompressible fluid, we decompose the total density $\varrho$ as $\varrho = \varrho_0 + \bar{\rho}(z) + \rho(x,z,t) $, where $\varrho_0$ is the reference density, $\bar\rho$ is the background density stratification as a function of depth and $\rho$ is the perturbation density. We consider the density changes from perturbations and stratification to be small compared to the reference density, so that $\bar\rho$, $\rho \ll \varrho_0$.

Given the triadic resonant condition in (\ref{eqn: resonant_condition}), it is easy to assume that the instability selects one particular triad, comprised of three distinct frequencies and wavenumbers for all time. More recently, our understanding of triad selection is evolving for finite-width beams. Indeed, while examining the transient start up of the instability, \cite{Koudella2006} note that not just one triad is responsible for the initial instability, rather, a number of triads form around the maximum linear growth rate. In addition, recent work by \cite{Fan2020} shows how classic TRI theory is unable to explain the instability in the context of a thin beam due to the broadband wavenumber spectrum corresponding to the primary beam. 

The novelty of the present paper lies in the examination of the long-term evolution of the instability. Due to the 11 m long tank used in the experimental set-up, we are able to observe the experiment for hours without interference from side wall reflections or significant changes to the stratification. We show experimentally that, over long time-scales, the constituent triadic waves synchronously modulate in amplitude and in the physical location of the two secondary wave beams. Further investigation shows that part of these modulations are coincident with the growth and decay of separate triads, all linked through the primary wave beam. Through two-dimensional weakly non-linear modelling, we are then able to show how the evolution of the instability in a finite-width beam is remarkably sensitive to these separate triads. This sensitivity is due to their affect on the residence time of the secondary wave beams with the underlying primary beam.  

The outline of the remainder of this paper is as follows. In $\S$ \ref{sec:exp_procedure} we detail the experimental set-up and processing procedure. In $\S$ \ref{sec: experimental results} we then present the experimental results, looking first at the initial observations in $\S$ \ref{sec: experimental processing} and then at the long-term evolution of the experiments $\S$ \ref{sec: long time development}. Based on these observations, we present the development of the two-dimensional weakly non-linear model in $\S$ \ref{sec: numerical set-up}. Here we outline the perturbation expansion used in $\S$ \ref{subsec: pert expansion} and the subsequent development of the numerical solution in $\S$ \ref{sec: model development} (comprising the two-dimensional advection scheme and the weakly non-linear interactions). In $\S$ \ref{sec: nonlinear interactions} we present the results of the model. We start with $\S$ \ref{subsec: nonlinear interactions}, where we examine the weakly non-linear interactions on their own before moving onto $\S$ \ref{subsec: non-linear model results}, where the results of the weakly non-linear two-dimensional model are given. Conclusions are then drawn in $\S$ \ref{sec: conclusions}.



\section{Experimental Procedure}
\label{sec:exp_procedure}

\subsection{Experimental setup}
\label{subsec: experimental setup} 


Experiments were undertaken in an 11 m long, 0.48 m deep, 0.29 m wide Perspex (acrylic) tank. Along a 1 m section of the tank floor, 2.5 m away from one end, sits the Arbitrary Spectrum Wave Maker (ASWaM), also known as the magic carpet. This flexible horizontal boundary can generate sinusoidal forcing \citep{Dobra2022, Dobra2021, Beckebanze2021} (as well as aperiodic configurations \citep{Dobra2019}), with the ability to vary amplitude, frequency and wavenumber in both the temporal and spatial domain. The wavemaker is comprised of a series of 96 computer-controlled linear actuators that sit below the tank. Each actuator is mounted to a vertical drive rod that passes through the base of the tank and connects to a 0.28 m long horizontal rod that spans the tank width. These rods are spaced at 10 mm intervals along the wavemaker. A 3 mm thick neoprene foam sheet covers the full length and width of the wavemaker thus interpolating between the horizontal rods to allow smooth forcing. The lengthwise edges of the neoprene slide against the tank walls and beneath there is a 80 mm cavity into which glycerol is added to help prevent salt crystallising and causing leakage around the seals that enable the drive rods to pass into the tank from the bank of actuators beneath. Provided the chosen waveforms preserve a zero-mean displacement across the length of the flexible surface, the pressure gradient available to drive flow around the edges of the neoprene foam is negligible. Thus flow in either direction between the cavity and the visualisation region may be considered negligible. When submerged in a stratified fluid, the wavemaker can generate quasi-two-dimensional, internal wave beams at amplitudes sufficient to permit wave breaking at distances away from the source. For full details of ASWaM's construction see \citet{Dobra2018} and \citet{Dobra2019}. 

The procedure for filling the tank is as follows. First, glycerol is gravity fed into the wavemaker cavity. The tank is then filled over the course of 8 hours with a linear salt-stratification using two computer controlled gear pumps, operated via the software DigiFlow \citep{Dalziel2007}. Each pump draws from either a fully saturated salt water or fresh-water reservoir. This filling method allows for more precise control of the density stratification compared with the traditional double bucket technique \citep{Oster1963}, enabling the density gradient and fluid depth to be pre-determined. The pumps used are Coleparmer Ismatec BVP-Z Analog gear pump drives mounted with two magnetically driven Coleparmer Micropump L20562 A-Mount Suction Shoe pump heads. The depth of the stratification $H$ = 0.45 $\pm$ 0.01 m. 

\begin{figure}
	{\includegraphics[width= 0.95\textwidth]{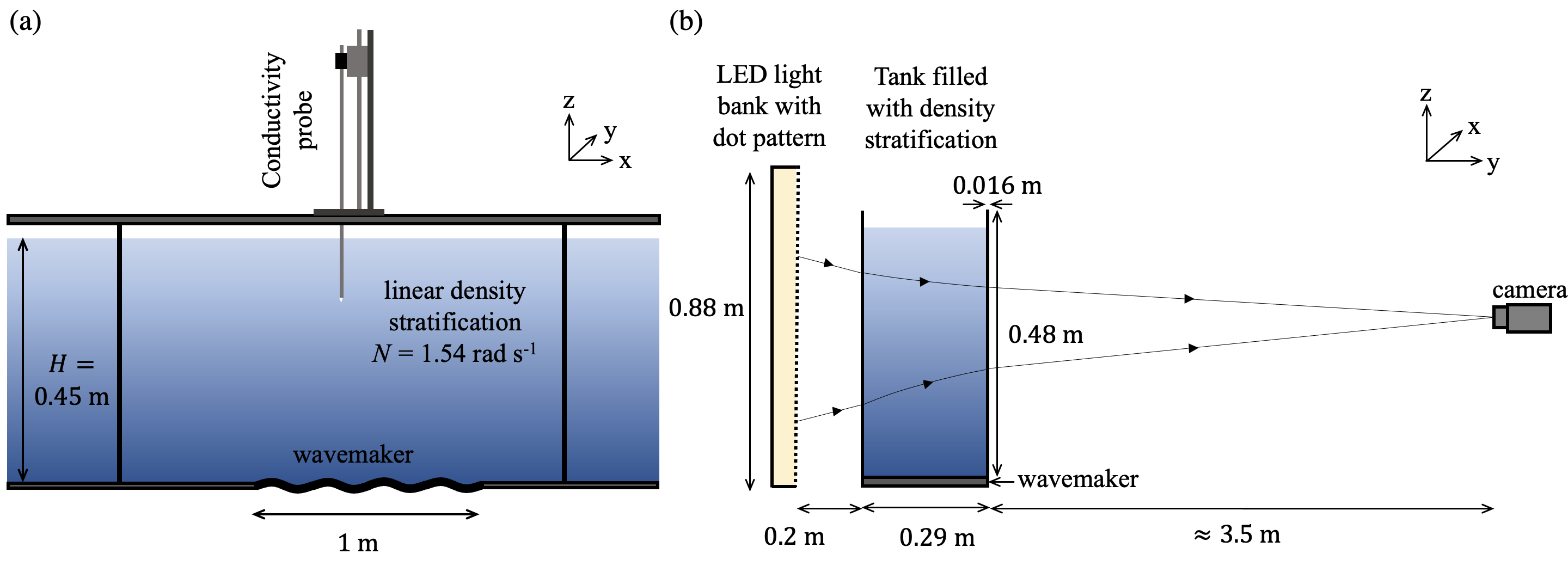}}
	\caption{(a) A schematic showing the front view of the tank as would be seen by the camera. The wavemaker is located along a 1 m section of the tank floor, sitting below a 0.45 m density stratification. A conductivity probe is mounted above the tank which is used to measure the density profile. (b) A schematic showing the side view of the tank in order to visualise the optical arrangement for Synthetic Schlieren. The thermal tunnel is not shown for clarity. }
	\label{fig: experimental setup}
\end{figure}

To measure the density profile created by the pumps, an aspirating conductivity probe is mounted to a linear traverse above the tank. One minute before the start and one minute after the end of an experiment, the probe is traversed downwards through the stratification to measure the conductivity of the saline solution passing through the probe tip. For the experimental campaign reported here, a linear density stratification with a buoyancy frequency $N$ = 1.54 $\pm$ 0.04 rad s$^{-1}$ is used. The variation in the buoyancy frequency is attributed to the evolution of the density stratification over the course of the week that the experiments were undertaken for. A schematic of the tank, as viewed from the front, can be seen in Figure \ref{fig: experimental setup}(a).

\subsection{Wave visualisation} 
\label{sec: experimental visulisation}

Synthetic Schlieren \citep{Dalziel2000, Dalziel2007} was used to visualise our experiments. This non-intrusive technique takes advantage of the differential refraction of light in a refractive index gradient and the Gladstone-Dale relationship between refractive index and fluid density, such that light rays curve towards regions of higher density. Internal waves make local perturbations to the density field and thus the direction of light rays passing through them will also be perturbed. The resulting distortion of a textured background image yields a measurable signal associated with the density perturbations. To minimise the effects of convective thermal fluctuations on the Synthetic Schlieren measurements in the air between the tank and the camera, a `thermal tunnel' ran from the camera lens to the perimeter of the visualisation region on the tank. 

A random dot pattern attached to an LED light bank was located 0.20 $\pm$ 0.04 m behind the tank, while a 12 MPixel ISVI IC-X12CXP camera with a Nikkor 35-135 mm zoom lens was located 3.50 $\pm$ 0.10 m from the front. This optical arrangement is shown in the side-view schematic in Figure \ref{fig: experimental setup}(b). The large distance between the camera and the tank was chosen in order to reduce parallax \citep{Thomas2009}. 

We compute the line-of-sight mean of the gradient vector of the density perturbation field $\rho$, which for convenience we non-dimensionalise according to
\begin{equation}
\boldsymbol{\beta} = (\beta_x, \beta_z) =  \frac{\lambda_{x_0}}{\varrho_0}\bigg(\frac{\partial{\rho}}{\partial x}, \frac{\partial{\rho}}{\partial z}\bigg),
\label{eqn: output from ss}
\end{equation}
where $\lambda_{x_0}$ is the horizontal wavelength of the primary wave beam given as \mbox{$\lambda_{x_0}$ = 2$\pi/|l_0|$}, where $l_0$ is the horizontal component of the primary wave vector $\boldsymbol{k}_0$. Our experiment is configured to generate and diagnose quasi-two-dimensional internal wave structures, up to the limit of wave breaking, the point at which the mapping of ray paths to density perturbations is no longer an aim.


\subsection{Internal wave forcing}

The experimental campaign presented in this paper comprises of 36 experiments. To reduce uncertainties associated with the test conditions both within the tank and in the laboratory ambient, the campaign was run within a seven-day period without refilling the tank, allowing a period of 3 hours between each experiment for any residual motion to dissipate. We note that other experimental campaigns were also undertaken over the course of a year that exhibited the same behaviour detailed below; for simplicity, we focus here on this one campaign. 

Following arguments laid out by \citet{Dobra2019}, for all the experiments in this campaign the vertical displacement $z = h(x,t)$ imposed on the neoprene foam to generate the primary beam, $\wave_0$, is 
\begin{equation}
\label{eqn: wavebeam envelope}
\hspace{-19mm} z = h(x,t) = \begin{cases}
\Re\big(f(t) \hspace{0.6mm} e^{i l_0 x} \cos^2\big(\frac{x-B}{8\pi^2}\big)\big), & \qquad \qquad  A < x < B,  \\
\Re\big(f(t) \hspace{0.6mm} e^{i l_0 x} \big),  & \qquad \qquad   B < x < C,   \\
\Re\big(f(t) \hspace{0.6mm} e^{i l_0 x} \cos^2\big(\frac{x-C}{8\pi^2}\big)\big), &  \qquad  \qquad C < x < D, \\
0, & \qquad \qquad   \textmd{elsewhere}, \\
\end{cases}
\end{equation}
where the locations $A, B, C, D$ are respectively $7\pi/|l_0|$, $9\pi/|l_0|$, $13\pi/|l_0|$, $15\pi/|l_0|$ and $l_0$ is set to $-$0.05 mm$^{-1}$, giving a horizontal wavelength of \mbox{$\lambda_{x_0}$ = 2$\pi/|l_0|$ $=$ 125.66 mm}. As we restrict $\omega_p > 0$ (for all $p$), having $l_0 < 0$ means the primary wave beam is propagating to the left. The spatial structure of the forcing, described by (\ref{eqn: wavebeam envelope}), takes the form of a beam with the inner two wavelengths reaching maximum amplitude and the outer wavelengths being smoothed by a cosine-squared envelope. Due to this cosine squared smoothing on the edges of the beam profile, we do not consider the full width, $D - A$, for energy transfer. Rather, we estimate the contribution from one of the smoothed edges using the integral measure employed by \citet{Dalziel1999}, giving a horizontal beam width of 
\begin{equation}
\label{eqn: delta width}
\Lambda_{x_0} = \Lambda_{0} / \cos\theta = 2\lambda_{x_0} + 2\int_{0}^{\lambda_{x_0}} \alpha \big(1 - \alpha\big) dx = 277.41 \textrm{ mm},
\end{equation}
where $\alpha = \cos^2(x/8\pi^2)$ is the smoothing function on the outer flanks of the beam profile. The temporal forcing $f(t)$ is then described as
\begin{equation}
\label{eqn: constant amp forcing}
f(t) = \begin{cases}
0, 	& \qquad  \qquad \hspace{4mm} t \leq 0 \: \mathrm{s}, \\ 
\eta_0 \hspace{0.6mm}\big(\frac{t}{30}\big)e^{-i\omega_0 t}, & \qquad \qquad  \hspace{4mm} 0 \leq t \leq 30 \: \mathrm{s}, \\
\eta_0 \hspace{0.6mm}e^{-i\omega_0 t}, & \qquad \qquad  \hspace{4mm}30 \: \mathrm{s} \leq t \leq t_\textrm{end} -30 \: \mathrm{s} \:, \\
\eta_0 \hspace{0.6mm}\big(\frac{t_\textrm{end} -t}{30}\big)e^{-i\omega_0 t}, & \qquad \qquad  \hspace{4mm} t_\textrm{end} -30 \: \mathrm{s} \leq t \leq t_\textrm{end} \:, 
\end{cases} \quad
\end{equation}
where $\omega_0$, $\eta_0$ and $t_\textrm{end}$ are respectively, the forcing frequency of 0.95 rad s$^{-1}$, the nominal forcing amplitude in mm of the primary beam and the end time of the experiment in seconds. Experiments are captured at 1 frame per second (fps), which is more than sufficient to capture the fast time evolution of the wave field given by the primary beam period $T_0 = 2\pi / \omega_0$.

The only two parameters to be varied in this experimental study are $t_\textrm{end}$ and $\eta_0$. The run time, $t_\textrm{end}$, is either 90 or 180 minutes, while the non-dimensional amplitude $\eta_0/\lambda_{x_0}$ ranges between 0.028 - 0.036. The amplitude threshold for the instability is achieved at $\eta_0/\lambda_{x_0} \approx$ 0.031 (reducing by $0.002$ throughout the week due a slow evolution of the stratification). Our focus is on the weakly non-linear regime, so we seek to minimise unnecessary mixing induced by wave actuation and limit our amplitudes to those just sufficient to exceed the instability threshold calculated by \citep{Davis1967}. 

Since the tank extends well beyond the field of view in both directions, internal wave beams with typical dominant wavenumbers of $\boldsymbol{k}_0 = (-0.05, -0.06)$ mm$^{-1}$ reflecting off the far boundary wall return to the viewing region with only 2$\%$ of their original amplitude, due to viscous dissipation over a beam length exceeding 4 m (the horizontal travel distance of the beam before re-interaction). We thus consider wave-wave interactions involving these reflected beams to be negligible. 

\section{Experimental results}
\label{sec: experimental results} 

\subsection{Initial observation and analysis} 
\label{sec: experimental processing} 

\begin{figure}
	\centering
	\begin{tikzpicture}
	\node at (1,1) {\includegraphics[width=1\textwidth]{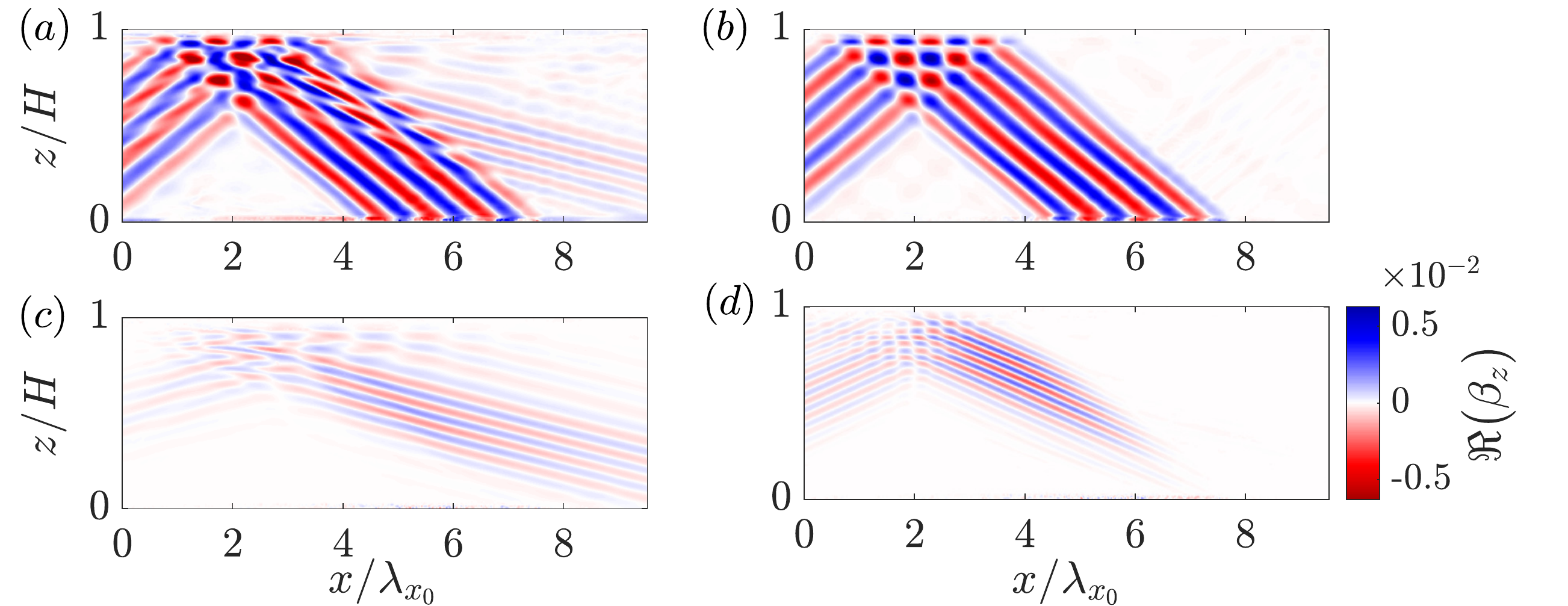}}; 
	\draw[black, thick, ->] (3, 2.5) -- (2.5, 2); 
	\draw[black, thick, ->] (-1.8, -0.3) -- (-1.65, 0.3); 
	\draw[black, thick, ->] (3.3, 0.3) -- (3, -0.3); 
	\draw[black, thick, -] (2.3, 2.6) -- (3.5, 2.6); 
	\draw[black, thick, -] (3.1, 1.9) -- (4.3, 1.9); 
	\draw[black, thick, -] (2.3, 2.6) -- (3.1, 1.9); 
	\draw[black, thick, -] (4.3, 1.9) -- (3.5, 2.6); 
    \node[black] at (2.3, 2.2) {\footnotesize $\boldsymbol{k}_0$};
    \node[black] at (-1.5, -0.25) {\footnotesize $\boldsymbol{k}_1$};
     \node[black] at (2.9, 0.05) {\footnotesize $\boldsymbol{k}_2$};
	\end{tikzpicture}
	\caption{(a) $\beta_z$ of the full flow field at $t/T_0 =$ 83 into an experiment forced at $\eta_0 /\lambda_{x_0}$ = 0.032. (b)--(d) The real part of three of the dominant frequencies produced from the DMD over $83 \leq t/T_0 \leq$ 86. The black arrows overlaid indicate the orientation of the respective wavenumber vectors $\boldsymbol{k}_p$. In panel (b) we see solely the wave field $\wave_0$ with $\omega_0/N$ = 0.62. In \mbox{(c) we see $\wave_1$ with $\omega_1/N =$ 0.23} and in (d) $\wave_2$ with \mbox{$\omega_2/N =$ 0.39}. The black box in (b) shows the spatial averaging domain $\langle \rangle_r$ used for the primary beam, discussed in $\S$ \ref{sec: model development}.}
	\label{fig: DMD TRI breakdown}
\end{figure}
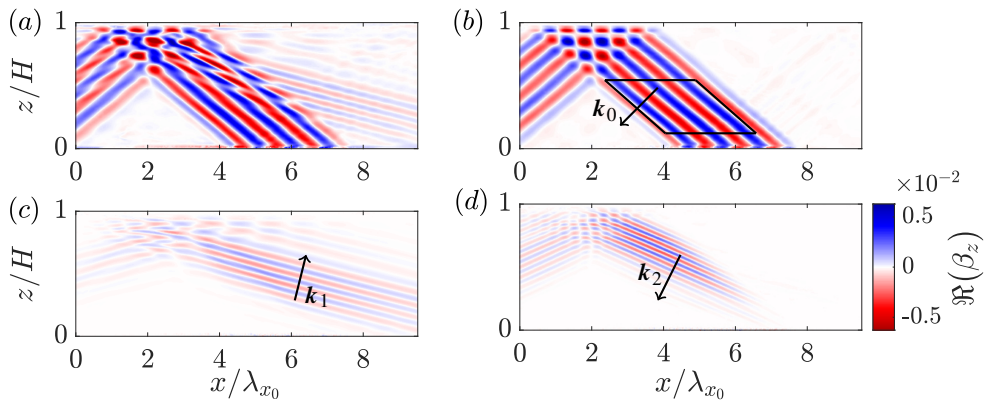

We start by examining one experiment from the set of 36 with an imposed amplitude displacement of $\eta_0 /\lambda_{x_0}$ = 0.032. Figure \ref{fig: DMD TRI breakdown}(a) shows $\beta_z$ over the visualisation region at \mbox{$t/T_0$ = 83}. Here, $\wave_0$, generated by the wavemaker, propagates energy up and to the left, at its respective group velocity $\boldsymbol{c}_{g_0}$. The group velocity is defined for all wave beams by 
\begin{equation}
\label{eqn: cg}
\boldsymbol{c}_{g_p} = \bigg(\frac{\partial}{\partial l_p}, \frac{\partial}{\partial m_p} \bigg) \omega_p = \textrm{sgn} (l_p) \frac{N m_p}{\kappa_p^3} (m_p, -l_p),
\end{equation}
where again the subscript $p = (0, 1, 2)$ corresponds to the primary beam and the two secondary beams, respectively, and the broadband wavenumber spectrum of each beam is approximated with a characteristic wavenumber. The primary beam $\wave_0$ reflects off the free surface, causing the vertical component of its group velocity to change sign and the wave-packets subsequently move down and to the left. An appropriate Reynolds number for the flow is given by $Re = \boldsymbol{c}_{g_0}/(\nu\kappa_0)$, where $\nu$ is the kinematic viscosity of 1 mm$^2$ s$^{-1}$. Here, $Re \approx 170$. As the selected input amplitude displacement of $\eta_0 /\lambda_{x_0}$ = 0.032 is above the instability threshold, $\wave_0$ becomes unstable, leading to the formation of two secondary beams. One of these beams, $\wave_1$, is clearly visible in Figure \ref{fig: DMD TRI breakdown}(a). This beam emanates from the central region of $\wave_0$ but moves in nearly the opposite direction, with a group velocity down and to the right. From Figure \ref{fig: DMD TRI breakdown}(a), the third beam, $\wave_2$, that completes the triad is not visible. In order to understand the underlying modal structure of these beams, the flow field $\boldsymbol{\beta}$, is decomposed using Dynamic Mode Decomposition (DMD). 


DMD works by preforming an eigen-decomposition of a linearised representation of the underlying evolution operator for a given flow field \citep{Schmid2010}. The `dynamic modes' are the recurrent spatial structures that accurately describe the dominant behaviour captured in the data sequence. Where DMD excels is in determining the frequencies and structure of the modes from short time series where there is a discrete spectrum that can reasonably be approximated by a combination of delta functions at slowly evolving frequencies. The ability to extract the modes from short time series allows exploration of the slowly evolving structure and frequency of the modes. This linear approximation for the evolution operator is valid for the experiments shown here due to the two discrete time-scales, whereby the slow time evolution of the beam amplitude is much less than the fast time-scales $\omega_p t$. 



The maximum number of dynamic modes is given by the number of input frames in the sequence $\delta t$ (in this case $\delta t$ = 20 s, as we use a frame rate of 1 fps, which is just greater than the slowest period of the triad $T_1$) and if the obtained mode is complex then it is coupled as a conjugate pair. Here, however, we are only interested in those modes with an eigenvalue modulus very close to one, as they represent the steady, non-decaying modes of the system. When instability occurs experimentally, four non-decaying modes are obtained, three of which are conjugate pairs of eigen-values. The real part of these three modes produced over the temporal window $83 \leq t/T_0 \leq$ 86 are given in Figure \ref{fig: DMD TRI breakdown}(b)-(d). As expected from the input forcing, Figure \ref{fig: DMD TRI breakdown}(b) corresponds to the input $\wave_0$ with non-dimensional frequency \mbox{$\omega_0/N$ = 0.62}. Figure \ref{fig: DMD TRI breakdown}(c) then corresponds to $\wave_1$ with \mbox{$\omega_1/N =$ 0.23} and (d) to the obscured $\wave_2$, which propagates with a group velocity up and to the left, with non-dimensional frequency \mbox{$\omega_2/N =$ 0.39}. To understand if these additional frequencies are the result of TRI, we sum the frequencies of the secondary beams and see that the temporal condition for triadic resonance, $\omega_0 = \omega_1 + \omega_2$, is satisfied. We remark that this frequency relationship is not enforced at any stage of experimental post-processing, but arises naturally from prominent signals found in the temporal spectrum. 




The fourth (non-decaying mode) corresponds to $\omega/N$ = 0 and is not shown here. This is generated from a two-wave interaction (TWI), in which two wave beams interact to produce a third wave beam, with a phase angle relationship
%
\begin{equation}
\label{eqn:  TWI resonant_condition}
\check{\phi} = \pm \phi_0 \mp \phi'_{0}.
\end{equation}
In this case, $\phi_0 = l_0 x + m_0 z - \omega_0 t $, corresponds to the phase angle of $\wave_0$ and $\phi'_{0} = l_0 x - m_0 z - \omega_0 t $ to its reflection, $\wave'_{0}$, from the free surface. These wave beams will sum to produce a third wave beam with wavenumber vector $\check{\boldsymbol{k}} = (0, 2m_0)$ aligned with the vertical and with zero frequency. This non-propagating disturbance can not be classed as a wave, but instead should be treated as a forced oscillatory structure that is confined to the interaction region of the primary beam with its reflection. If considered analytically \citep{Thorpe1987} or numerically \citep{Grisouarda2013} in a two-dimensional setting, only weak horizontal vorticity is generated, which is partially suppressed by the background stratification \citep{Beckebanze2019}. When considered in a three-dimensional setting, however, \citet{Grisouarda2013} show, both experimentally and numerically, how a stronger slowly evolving three-dimensional horizontal mean flow develops from the interaction region of the primary beam with its reflection. This flow has a vertical component to its vorticity field. Indeed, if viscous attenuation and cross-beam variations are present, it is possible for a three-dimensional mean flow to be generated from the wave beam interacting with itself, as shown analytically by \citet{Kataoka2015} and experimentally by \citet{Bordes2012b}. In all of the three-dimensional cases cited above, however, the wave beam is propagating in a tank wider than the beam width. This allows for a recirculating mean flow to develop in the transverse direction, outside of the spatial extent of the beam. As noted by \citet{Sutherland2006} in experiments where wave beams are confined laterally by tank side walls, as is the case in the experiments presented here, horizontal mean flow of this type is unable to develop. The observed zero-frequency mode in our experiments, closely resembles the two-dimensional simulations of \cite{Grisouarda2013} and, while the disturbance does slowly exit the interaction region of $\wave_0$ and $\wave'_{0}$, no strong recirculating mean flow is seen to develop and as such does not impact the evolution of TRI described here.


%

\begin{figure}
	\centering
	\begin{tikzpicture}
	\node at (1,1) {\includegraphics[width=0.6\textwidth]{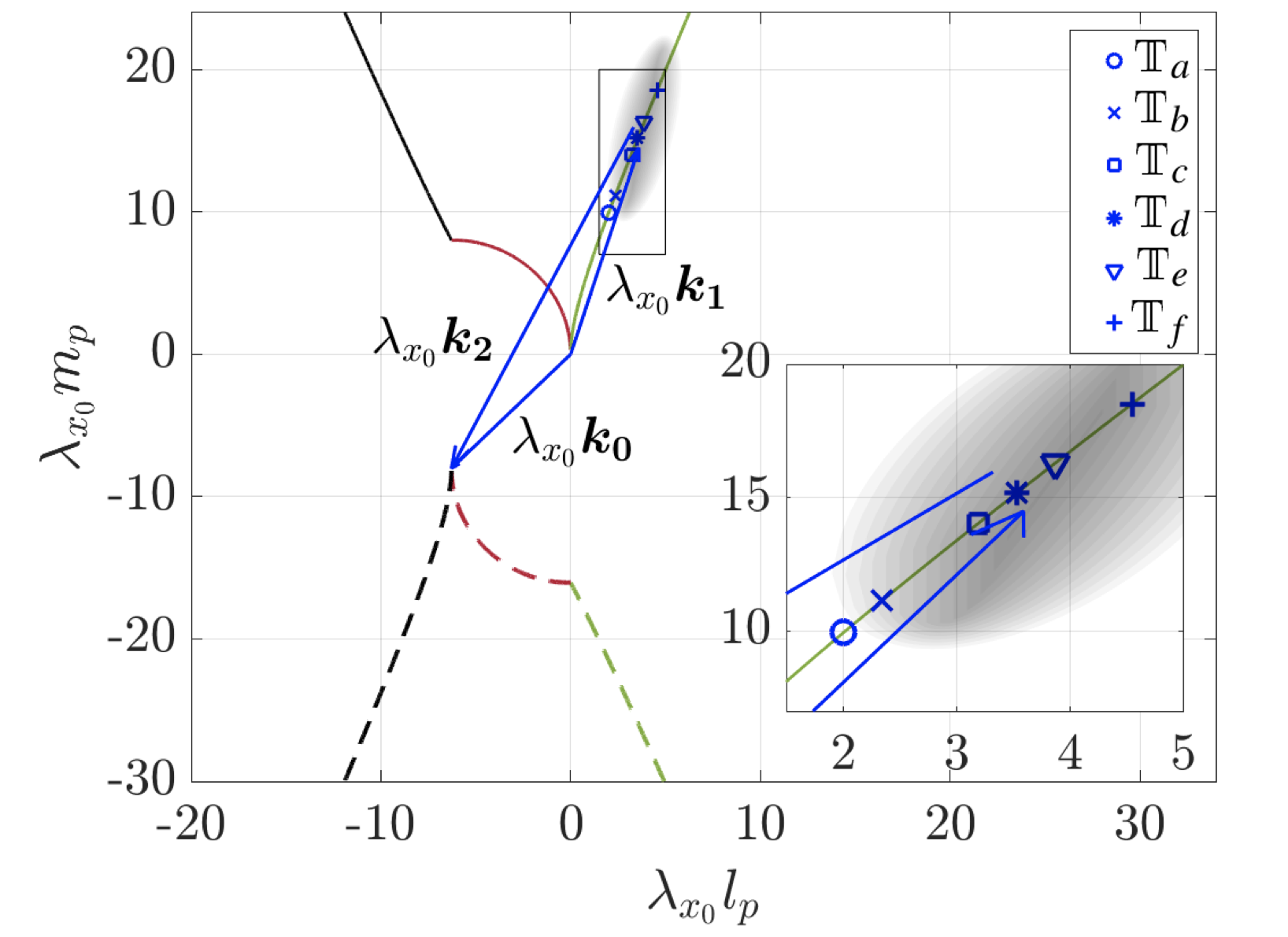}} ; 
	\end{tikzpicture}
	\caption{The underlying solid and dashed black, red and green curves give all of the possible locations for the tip of $\boldsymbol{k}_1$ that satisfy both the dispersion relationship (\ref{eqn: dispersion}) and the TRI condition (\ref{eqn: resonant_condition}) for a given $\wave_0$. The dark blue arrows show the experimentally produced, characteristic, wavenumber vectors of the resonant triad shown in Figure \ref{fig: DMD TRI breakdown}, obtained from taking the Fourier transform in $(x,z)$ of the gradient field. The shaded grey region then indicates the range of wavenumber vectors obtained over the course of the experiment. The six dark blue marks correspond to the different triad wave vector configurations used in the weakly non-linear modelling and are discussed in $\S$ \ref{subsec: nonlinear interactions}. The panel in the bottom right corner shows an enlarged view of the region enclosed by the black rectangle.}
	\label{fig: loci_curve}
\end{figure}

We proceed to determine the wave vectors corresponding to the primary and secondary wave beams by taking our frequency-decomposed gradient field over the temporal window $83 \leq t/T_0 \leq$ 86 -- the real parts of which are shown in Figures \ref{fig: DMD TRI breakdown}(b)–(d) -- and calculating a two-dimensional power spectra on each constituent field separately. Each image is embedded in a zero filled matrix in order to improve resolution and limit spatial aliasing. The wavenumber is determined by fitting a quadratic curve to the peak of the resultant power spectra and finding the wavenumber corresponding to the peak of the curve. This procedure is preformed on every row and column of the domain and subsequently mean averaged over both spatial dimensions. The smallest resolvable length scale is 2 pixels, equivalent to the non-dimensional length $x/\lambda_{x_0} = 0.005$, given by the ratio of pixel resolution to region size. As the analysis preformed on the horizontal density gradient $\beta_x$ provides similar results to that of the vertical $\beta_z$, we use only the results from the vertical gradient for simplicity. The non-dimensional characteristic wavenumbers for the vertical gradient fields shown in Figure \ref{fig: DMD TRI breakdown} are $\lambda_{x_0}\boldsymbol{k}_0$ = ($-$6.28, $-$8.29), $\lambda_{x_0}\boldsymbol{k}_1$ = (3.73, 14.60), and $\lambda_{x_0}\boldsymbol{k}_2$ = ($-$9.93, $-$24.88). These wave vectors are shown by the blue arrows on Figure \ref{fig: loci_curve}, where the underlying black, red and green curves provides the locus of all possible solutions for $\boldsymbol{k}_1$ given $\boldsymbol{k}_0$, based on both the dispersion relationship (\ref{eqn: dispersion}) and the TRI condition (\ref{eqn: resonant_condition}). 

While the calculated characteristic wave vectors shown in Figure \ref{fig: loci_curve} lie almost in a closed triangle, their alignment is not perfect, potentially indicating that the spatial triadic resonance condition $\boldsymbol{k}_0 = \boldsymbol{k}_1 + \boldsymbol{k}_2$ is not exactly satisfied. The reason for this slight misalignment is due to three factors. Firstly, there is the impact of inevitable experimental noise. Secondly, as we are considering finite-width beams as opposed to plane-waves, each beam is comprised of a broadband wavenumber spectrum. By defining a single characteristic wavenumber for the beam -- taken from the peak of the Fourier spectrum -- we are therefore approximating this wavenumber distribution. Thirdly, we are assuming that the spatial structures of $\wave_1$ and $\wave_2$ are uniform over the field of view. In fact, as the experiment progresses, significant modulations to the structures of $\wave_1$ and $\wave_2$ are observed, revealing that this assumption of spatial uniformity is inappropriate.


%

While it is clear that TRI was indeed being witnessed experimentally in a finite-width beam, this in itself is not novel. In an experiment actuated by an oscillating cylinder, \citet{Clark2010} attribute the breakdown of a wave beam due to TRI, showing how the instability evolves from infinitesimal perturbations in the flow. This work has recently been developed by \citet{Fan2020}, who discuss the validity of TRI theory in thin wave beams. Moreover \citet{Joubaud2012} and \citet{Bourget2013} clearly show the growth of the instability for a finite-width beam in experiments using their sidewall wavemaker. In our work, the regime of interest is not the initial growth of the instability, but rather the finite-amplitude unsteady modulations that occur afterwards. As noted, the expected saturated equilibrium state for the weakly non-linear instability is not observed, rather we witness slow modulations of the amplitudes and structures of the constituent beams in the triad, revealing much more dynamical behaviour than anticipated. We investigate the long-term evolution of this unsteady behaviour for the remainder of the paper.

\subsection{Long-time development} 
\label{sec: long time development}

\begin{figure}
	\centering
	\begin{tikzpicture}
	\node at (1,1) {\includegraphics[width=1\textwidth]{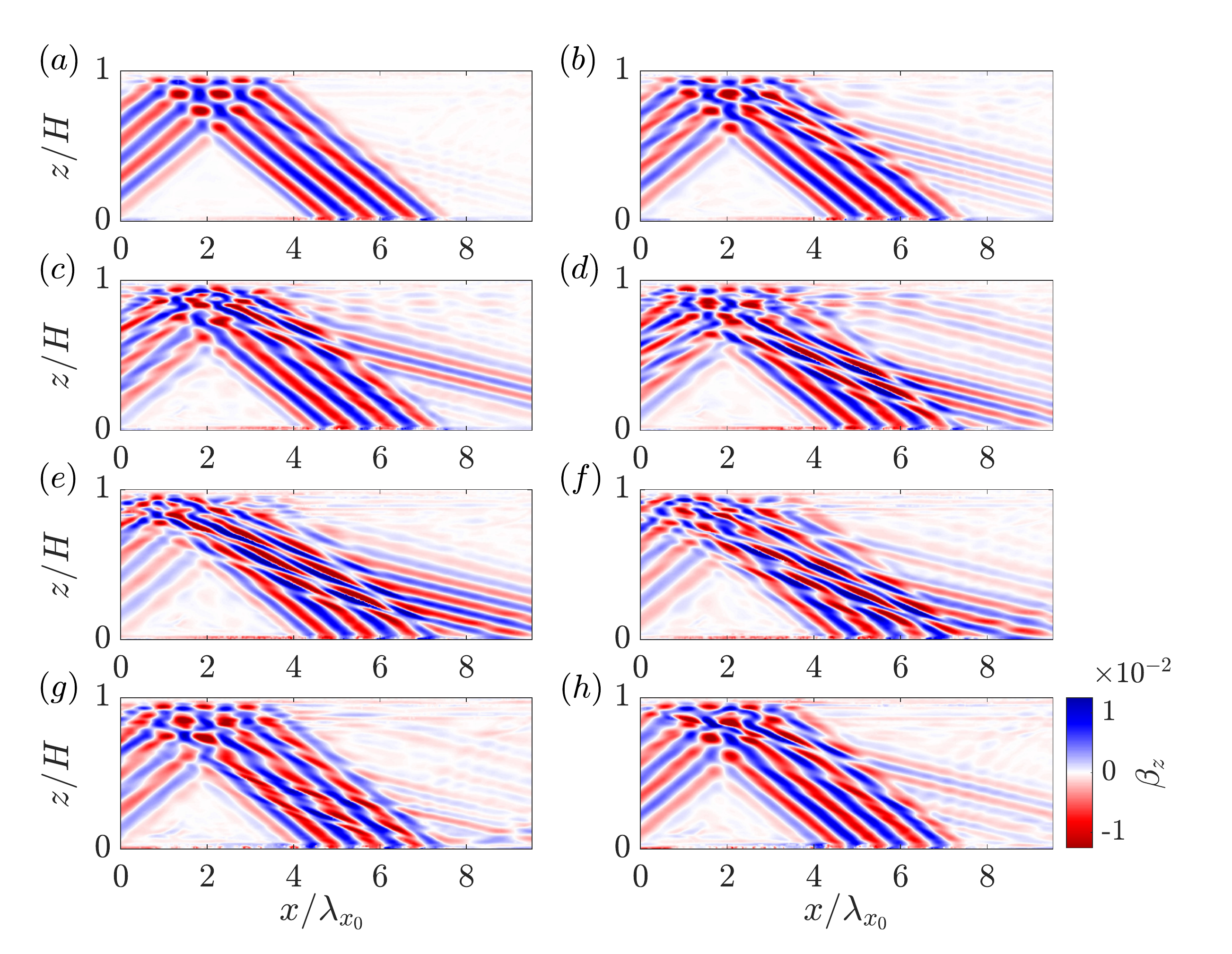}}; 
	\draw[black, thick, -] (0.1, -2.4) -- (-2.15, -1.9) ; 
	\draw[black, thick, -] (-1.4, -2.7) -- (0.1, -3.0); 
	\end{tikzpicture}
	\caption{Sequence of images showing the vertical gradient of the density perturbation of an experiment with $\eta_0 /\lambda_{x_0}$ = 0.032. (a) $t/T_0$ = 53, (b) $t/T_0$ = 83, (c) $t/T_0$ = 113, (d) $t/T_0$ = 144, (e) $t/T_0$ = 174, (f) $t/T_0$ = 213, (g) $t/T_0$ = 432, (h) $t/T_0$ = 582. The black lines in (g) indicate where the $\wave_1$ beam changes frequency, evidenced by the subtle change in angle in between the two lines.}
	\label{fig: time_snaps}
\end{figure}

Figure \ref{fig: time_snaps} shows 8 instantaneous images of the experiment shown in Figure \ref{fig: DMD TRI breakdown}. Figure \ref{fig: time_snaps}(a) is captured at $t/T_0$ = 53 into the experiment, just as $\wave_0$ becomes visibly unstable. By the instant shown in \ref{fig: time_snaps}(b) (the same image shown in Figure \ref{fig: DMD TRI breakdown}(a)) $\wave_1$ has clearly developed, with a group velocity propagating down and to the right. A particularly interesting feature of the subsequent time frames is the modulation of $\wave_1$ over time. Not only is its region of generation not constant -- it migrates across the full height of $\wave_0$ -- the beam itself also varies in both intensity and width. This migratory behaviour persists for the full duration of the experiment, which lasts for over 800 periods of the primary beam.

%


Further quantitative analysis of this peculiar behaviour requires us to calculate the amplitude of the individual resonant wave beams $\wave_0$, $\wave_1$ and $\wave_2$. Decomposing by frequency into complex constituent fields using DMD, we find the inverse gradient (potential) field $\rho_p/\varrho_0$ by integrating both the real and imaginary components of $\boldsymbol{\beta}$. We then isolate the wave beam of interest further using a Hilbert Transform, first used for internal waves by \citet{Mercier2008}. This filtering technique is applied to isolate the quadrant of Fourier space containing the wave vectors of the beam of interest from other signals of the same temporal frequency e.g. separating $\wave_0$ from its reflection from the free surface, $\wave'_0$. 

In order to have a singular value for amplitude that is independent of space, we then spatially mean average $\rho_p/\varrho_0$ over the whole field of view denoted $\langle \rangle_w$. This choice of spatial averaging ensures that our measure of amplitude is decorrelated with the position of a beam in space, a topic that will warrant further discussion in $\S$ \ref{subsec: nonlinear interactions}. An unavoidable consequence of this choice, however, is that this average measure no longer represents the local amplitude within a beam. To account for this, the other region used for spatial averaging is shown by the black box in Figure \ref{fig: DMD TRI breakdown}(b), which we donate as $\langle \rangle_r$. This region is only ever used for the primary beam and is used to compare the experimental input amplitude with the two-dimensional and zero-dimensional modelling discussed in $\S$ \ref{sec: model development}.

We then infer the amplitude from the measured displacements of the scalar two-dimensional stream function $\Psi=\Psi(\boldsymbol{x},t)$ for each field, where the velocity vector $\boldsymbol{u} = \grad \times (\Psi \hat{\boldsymbol{y}})$ and $\grad = (\partial /\partial x, \partial /\partial z)$. Assuming an oscillatory form $\Psi_p =\tilde{\Psi}_p e^{i\phi_p}$ for each triadic wave beam, we define our spatially averaged amplitude for each constituent field as $\langle\tilde{\Psi}_p\rangle_w$, a quantity that is independent of the oscillatory time-scale. As our focus here is on the slow time-scale evolution of the field, we allow $\langle\tilde{\Psi}_p\rangle_w = \langle\tilde{\Psi}_p(t)\rangle_w$, on a slow time-scale well-separated from the oscillation period of all wave beams in the system. In the same way, we define the density perturbation $\rho_p = \tilde{\rho}_pe^{i\phi_p}$ for each field. 

Substituting the above forms for stream function and density into the inviscid linear conservation of mass equation $\partial \rho/\partial t = - w (d\bar{\rho}/d z)$, and cancelling the fast time-scales, we find the reduced stream function amplitude for each wave field using 
%
\begin{equation}
\label{eqn: psi from SS} 
\langle|\tilde{\Psi}_p|\rangle_w = \bigg|\frac{\omega_p}{l_p} \frac{g}{N^2}\bigg|\frac{\langle|\tilde{\rho}_p|\rangle_w}{\varrho_0}, 
\end{equation}
where $\omega_p$ and $l_p$ are the frequency and horizontal wavenumber of the given beam $p$, and $\langle|\tilde{\rho}_p|\rangle_w$ is the root mean square (magnitude) of the complex output of the wave field after being spatially filtered by the Hilbert Transform. 

\begin{figure}
	\centering
	\includegraphics[width=1\textwidth]{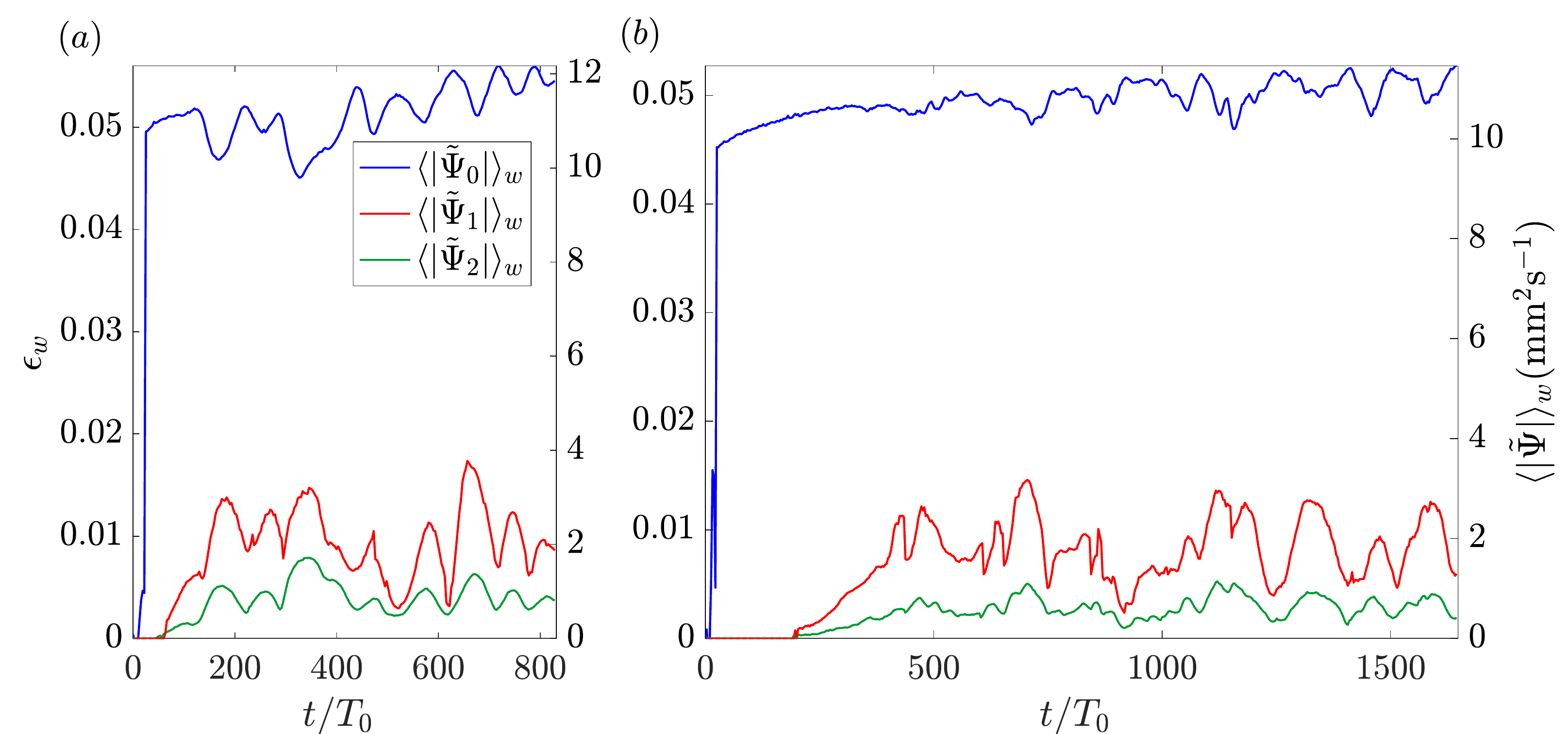}
	\caption{The non-dimensional amplitude of the reduced stream function $ \epsilon_w = \kappa_0^2\langle|\tilde{\Psi}_p|\rangle_w/N$, for $\wave_0$ (blue), $\wave_1$ (red), $\wave_2$ (green) for two experiments. Here $\langle|\tilde{\Psi}|\rangle_w$ is defined in (\ref{eqn: psi from SS}), where the spatial averaging of each signal is performed over the whole domain $\langle \rangle_w$. The dimensional amplitude of each wave field is given on the right axis. (a) $t_{\textrm{end}}/T_0 = 816$, $\eta_0 /\lambda_{x_0}$ = 0.032. (b) $t_{\textrm{end}}/T_0 = 1633$, $\eta_0 /\lambda_{x_0}$ = 0.032.}
	\label{fig: amplitude modulation}
\end{figure}

Figure \ref{fig: amplitude modulation} shows the amplitude calculated from (\ref{eqn: psi from SS}) for two experiments. In (a) we show the same experiment as Figure \ref{fig: time_snaps}, while (b) corresponds to another experiment with the same amplitude ($\eta_0 /\lambda_{x_0}$ = 0.032) but with a much longer run time ($t_{\textrm{end}}/T_0 = 1633$). We first note that the growth of the secondary wave beams appears earlier in (a) than (b) and that the maximum amplitude of the primary wave beam in (a) is larger, despite both experiments having the same amplitude displacement, $\eta_0 /\lambda_{x_0}$, from the wavemaker. This is due to the deterioration of the stratification throughout the week of experiments, which results in decreased transmission from the wavemaker to $\wave_0$ and hence a slight reduction in the instability threshold. This change emphasises the need to use the measured wave beam amplitude, calculated using (\ref{eqn: psi from SS}), as opposed to the imposed displacement from the wavemaker. 

Another observable feature in Figure \ref{fig: amplitude modulation} is the gradual increase of the mean amplitude of $\wave_0$ over time. This behaviour is also seen in lower amplitude forcing experiments that did not become unstable to TRI (not shown here). This increase can not be directly due to the instability, as the TRI mechanism transfers energy from the primary beam to the two secondary wave beams, as opposed to injecting energy into the primary wave beam. Rather, the amplitude increase is believed to be due to the peristaltic motion of the wavemaker leading to a sharpening of the stratification directly above the wavemaker, resulting in an increased transmission efficiency between the energy transfer from the wavemaker to the internal waves.

The most prominent feature in Figure \ref{fig: amplitude modulation} is the amplitude modulations of all the triadic wave beams, observed in every experiment that became unstable. While these modulations were anticipated from qualitatively observing the experiments, quantitatively they are found to be unexpectedly large and without obvious periodicity. This behaviour was so striking that we initially sought explanations unrelated to the physics of the system, such as measurement errors in converting raw video footage to density gradient fields or discrepancies that might be introduced by frequency-decomposition into constituent fields. After careful examination of both the raw data and the tool chain, including replicating the harmonic analysis of \cite{Mercier2008} -- a technique that relies solely on Fourier transforms to isolate waves before calculating $\tilde{\Psi}$ using (\ref{eqn: psi from SS}) -- we were able to discount all extraneous sources that could contribute to these structural modulations.

In Figure \ref{fig: amplitude modulation}, the amplitudes of $\wave_1$ (red) and $\wave_2$ (green) are positively correlated; their amplitudes are almost scaled values of each other. Meanwhile, the amplitude of $\wave_0$ is negatively correlated with $\wave_1$ and $\wave_2$. When $\wave_0$ is at a local maximum, the amplitudes of $\wave_1$ and $\wave_2$ are concurrently at a local minimum and then versa when the amplitude of $\wave_0$ is at a minimum. This coupling of the modulations in amplitude between $\wave_0$ and the secondary $\wave_1$ and $\wave_2$, reveals a continuous energy exchange flux between the wave beams in the triad that does not saturate to a steady equilibrium. For these experiments, the pattern of slow modulation appears to be independent of the primary wave beam amplitude, as, when normalised, the amplitude ratios $\wave_1/\wave_0$ and $\wave_2/\wave_0$ are similar across all experiments that become unstable independent of the amplitude of the forcing. Despite the clear pattern of modulations shown in Figure \ref{fig: amplitude modulation}, there is sufficient randomness that the signal does not have a clear dominant frequency in Fourier space. This observation is common to all experiments where instability develops. 

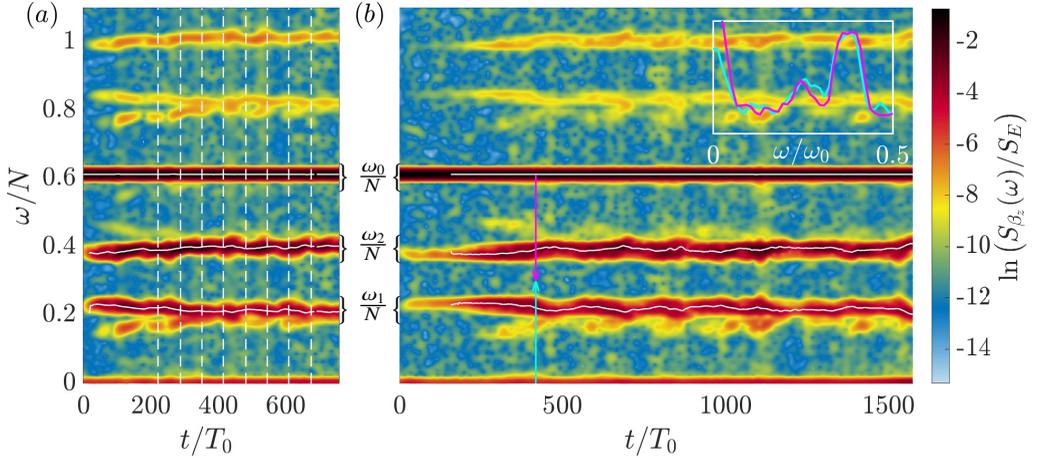
\begin{figure}
	\centering
	\begin{tikzpicture}
	\node at (1,1) {\includegraphics[width=1\textwidth]{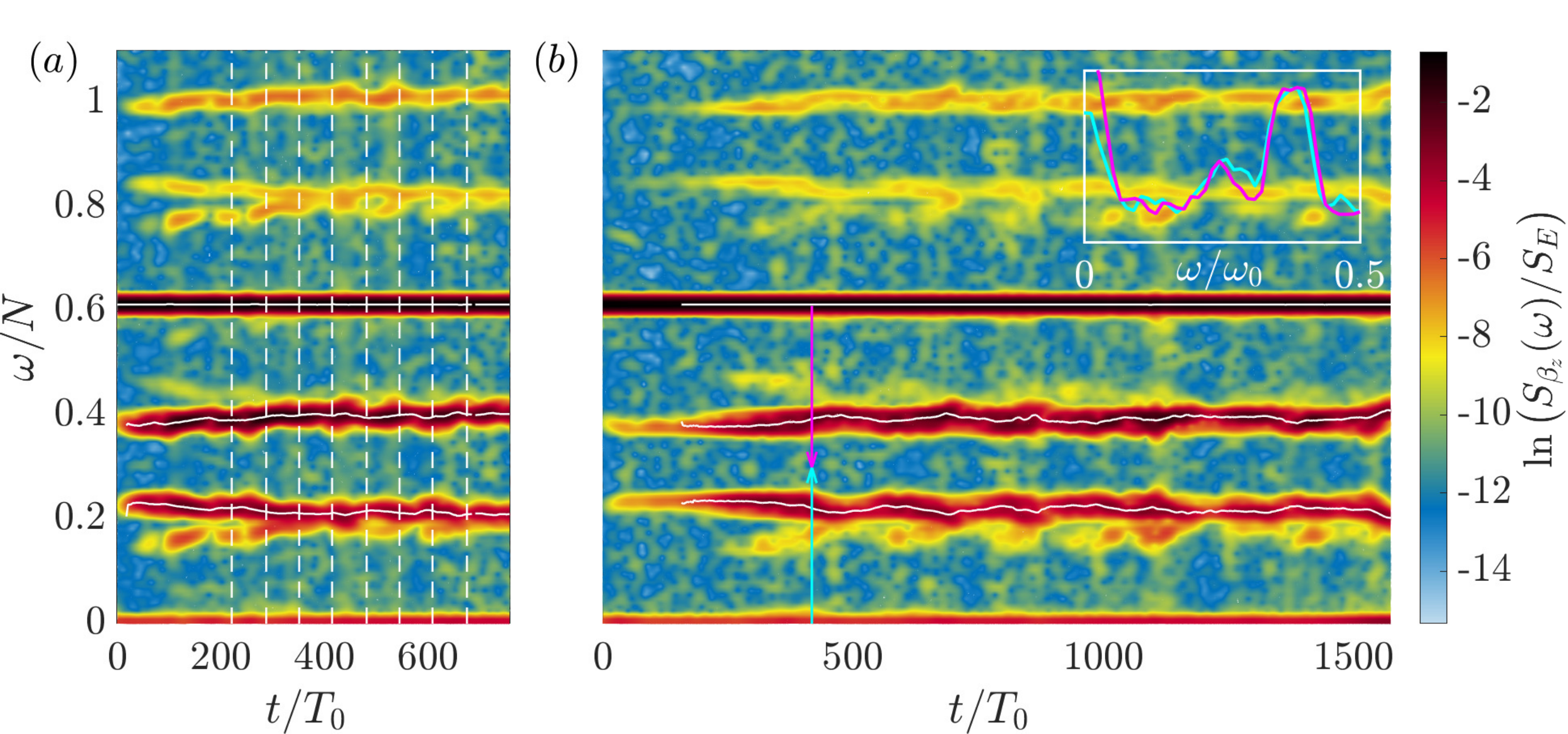}}; 
	\node[black] at (-0.91, 0.6) {\footnotesize $\frac{\omega_2}{N}$};
	\node[black] at (-0.91, -0.2) {\footnotesize $\frac{\omega_1}{N}$};
	\node[black] at (-0.91, 1.5) {\footnotesize $\frac{\omega_0}{N}$};
	\node[black] at (-1.3, 0.55) {\footnotesize $\big\}$};
	\node[black] at (-1.3, -0.25) {\footnotesize $\big\}$};
	\node[black] at (-1.3, 1.5) {\footnotesize $\big\}$};
	\node[black] at (-0.6, 0.55) {\footnotesize $\big\{$};	
	\node[black] at (-0.6, -0.25) {\footnotesize $\big\{$};
	\node[black] at (-0.6, 1.5) {\footnotesize $\big\{$};	

	\end{tikzpicture}
	\caption{Time-frequency spectra computed for the experiments in Figure \ref{fig: amplitude modulation}. The spectral density is computed by (\ref{eqn: spectrogram}) and is normalised by the total energy $S_E$ = $\sum_{z=0}^{H} S_{\beta_{z}}(\omega)^2$ for each instant in time. The dominant frequencies for each experiment obtained from the DMD frequency-decomposition are overlaid in white. The white dashed lines in (a) indicate the times of the instantaneous images shown in Figure \ref{fig: wavenumber spectrograms}. The subplot overlaid on (b) shows a transect in time at $t/T_0$ = 452, marked by the black and magenta arrows. Here we have plotted $\ln (S_{\beta_z} (\omega)/S_E)$ in cyan and $\ln (S_{\beta_z} (\omega_0 - \omega)/S_E)$ in magenta against $\omega/\omega_0$. }
	\label{fig: spectrograms}
\end{figure}

Both the physical positioning of the secondary wave beams (seen in Figure \ref{fig: time_snaps}) and their amplitudes (shown in Figure \ref{fig: amplitude modulation}) undergo slow modulation. Less obvious is that the beam frequencies also simultaneously modulate. This is evidenced in Figure \ref{fig: time_snaps}(g), where the angle of $\wave_1$ is noticeably closer to the horizontal in the lower part of the domain in comparison to the upper part of the domain (in between the two black lines). To further understand the slow evolution of beam frequencies, Figure \ref{fig: spectrograms} shows the temporal-frequency spectra computed using a Fourier-transform for both experiments presented in Figure \ref{fig: amplitude modulation}, along with the corresponding DMD estimates of the triadic frequencies overlaid in white. The amplitude of the spectra is determined by 
%
\begin{equation}
\label{eqn: spectrogram} 
S_{\beta_{z}}(\omega, t) = \bigg\langle\bigg| \frac{1}{T_T}\int_{-\infty}^{+\infty}\beta_{z}(x,z,t') e^{-i2\pi (\omega t')} W(t'-t; T_T) dt' \bigg|^2\bigg\rangle_{w},
\end{equation}
where $W(t'; T_T)$ is a Hamming window of non-dimensional width $T_T/T_0$ = 39. For the frame rate of 1 fps, the highest resolvable frequency (shortest time period) is $\omega/N = 4.08$. Several windowing functions were tested, and were not found to significantly affect the spectrogram results. The angled brackets, $\langle \rangle_{w}$, again indicate that the results are averaged across the whole visualisation region. This underlying spectrogram, calculated using (\ref{eqn: spectrogram}), therefore, reveals the details about the distribution of the frequency spectra for $\omega_1$ and $\omega_2$. In contrast, as we are only selecting the three dominant modes obtained from the DMD over short time intervals ($\delta t/T_0 = 3$), this methods approximates the underlying energy spectrum by a series of delta functions, allowing us to clearly see the slow-time evolution of these dominant modes.

Both spectrograms in Figures \ref{fig: spectrograms}(a) and (b) show a clear peak at $\omega_0/N$ = 0.62 for all time, consistent with the imposed displacement from ASWaM. Both secondary beams emerging from the instability become visible at approximately $t/T_0$ = 50, with peaks in the spectra around $\omega_1/N$ $\approx$ 0.23 and $\omega_2/N$ $\approx$ 0.39, though subsequently these modulate on a slow time-scale throughout the duration of an experiment. The overlaid DMD frequency estimates match almost perfectly the three frequency peaks on the spectrogram, following the same pattern of slow modulations. Despite this modulation, the temporal triadic relationship $\omega_0 = \omega_1 + \omega_2$, is satisfied at all times for the frequencies obtained from the DMD. As noted previously, the triadic requirement is not built into the DMD analysis. Interestingly, a similar variation in frequency has been witnessed by both \citet{Bourget2013} and \citet{Brouzet2016} in their experimental studies, however the phenomenon was not the focus of their work.

In addition to the triadic frequencies, there are three other distinct frequency bands found in the time-frequency spectrograph. The band with the lowest frequency corresponds to $\omega/N$ $\approx$ 0, which was also observed from the DMD and has already been discussed. The other two frequencies \mbox{$\omega/N$ $\approx$ 0.84} and $\omega/N$ $\approx$ 1 correspond to two different TWIs, given in (\ref{eqn:  TWI resonant_condition}), between $\wave_0$ and either $\wave_1$ or $\wave_2$, respectively. 

\begin{figure}
	\centering
	\begin{tikzpicture}
	\node at (1,1) {\includegraphics[width=1\textwidth]{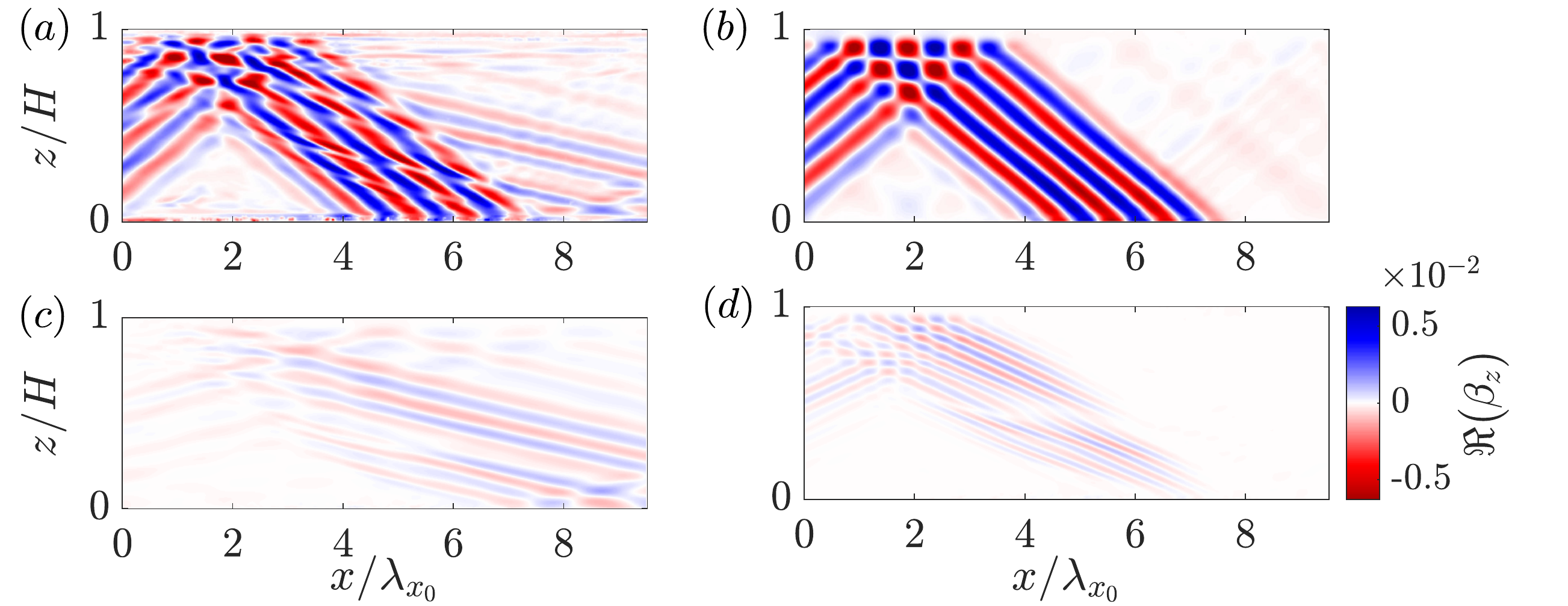}}; 
	\draw[black, dashed] (2.5,-0.15) -- (4.1,-0.15) -- (4.1,0.25) -- (2.5,0.25) -- (2.5,-0.15);
	\end{tikzpicture}
	\caption{(a) $\beta_z$ of the density perturbation field for the experiment presented in Figure \ref{fig: time_snaps} at $t/T_0$ = 481. (b)–(d) The real part of the three pairs of dominant modes extracted using DMD over a time-window from 481 $\leq t/T_0 \leq$ 483. Specifically, (b) $\wave_0$ with $\omega_0/N$= 0.617 (c) $\wave_1$ with $\omega_1/N$ = 0.211 (d) $\wave_2$ with $\omega_2/N$ = 0.405. The black dashed box in (d) indicates the region of discontinuity in the $\wave_2$ beam.}
	\label{fig: DMD messy}
\end{figure}

What is perhaps most striking from these time-frequency spectra is how, at certain points in time, there are multiple sets of $\wave_1$ and $\wave_2$ associated with the instability. This is shown by the convergent `wisps' on the $\omega_1/N$ and  $\omega_2/N$ bands, where additional secondary beam pairs appear and merge with the continuous mode. This is highlighted for $t/T_0$ = 452 by the inset in Figure \ref{fig: spectrograms}(b). Here we have plotted $\ln (S_{\beta_z} (\omega)/S_E)$ in cyan and $\ln (S_{\beta_z} (\omega_0 - \omega)/S_E)$ in magenta against $\omega/\omega_0$. The presence of a spectrum of triadic relations here is evidenced by the strong correlation between the two traces, indicating that the triadic requirement $\omega_1 + \omega_2 = \omega_0$ persists across all the spectrum. To analyse these frequency modulations further, Figure \ref{fig: DMD messy} shows the real part of the dynamic modes associated with the three dominant pairs of frequencies from the DMD over frames 481 $\leq t/T_0 \leq$ 483, from the experiment presented in Figure \ref{fig: time_snaps}. Unlike its earlier counterpart in Figures \ref{fig: DMD TRI breakdown}(c) and (d), where there was one distinct frequency and wavenumber pair for both $\wave_1$ and $\wave_2$, Figures \ref{fig: DMD messy}(c) and (d) show that, at this instant in time, TRI is occurring at two different locations over the height of the primary beam. For both of these modes, the signal is discontinuous across a transition region where the two out-of-phase wave beams de-constructively meet, highlighted by the black dashed rectangle in Figure \ref{fig: DMD messy}(d). Indeed, the presence of these separate beams is confirmed by splitting the domain in half horizontally and preforming the DMD analysis separately on the two halves. For the upper half of the domain $\omega_1/N$ = 0.211 and $\omega_2/N$ = 0.406, while in the bottom half of the domain $\omega_1/N$ = 0.214 and $\omega_2/N$ = 0.403. 





\begin{figure}
	\centering
	\includegraphics[width=1\textwidth]{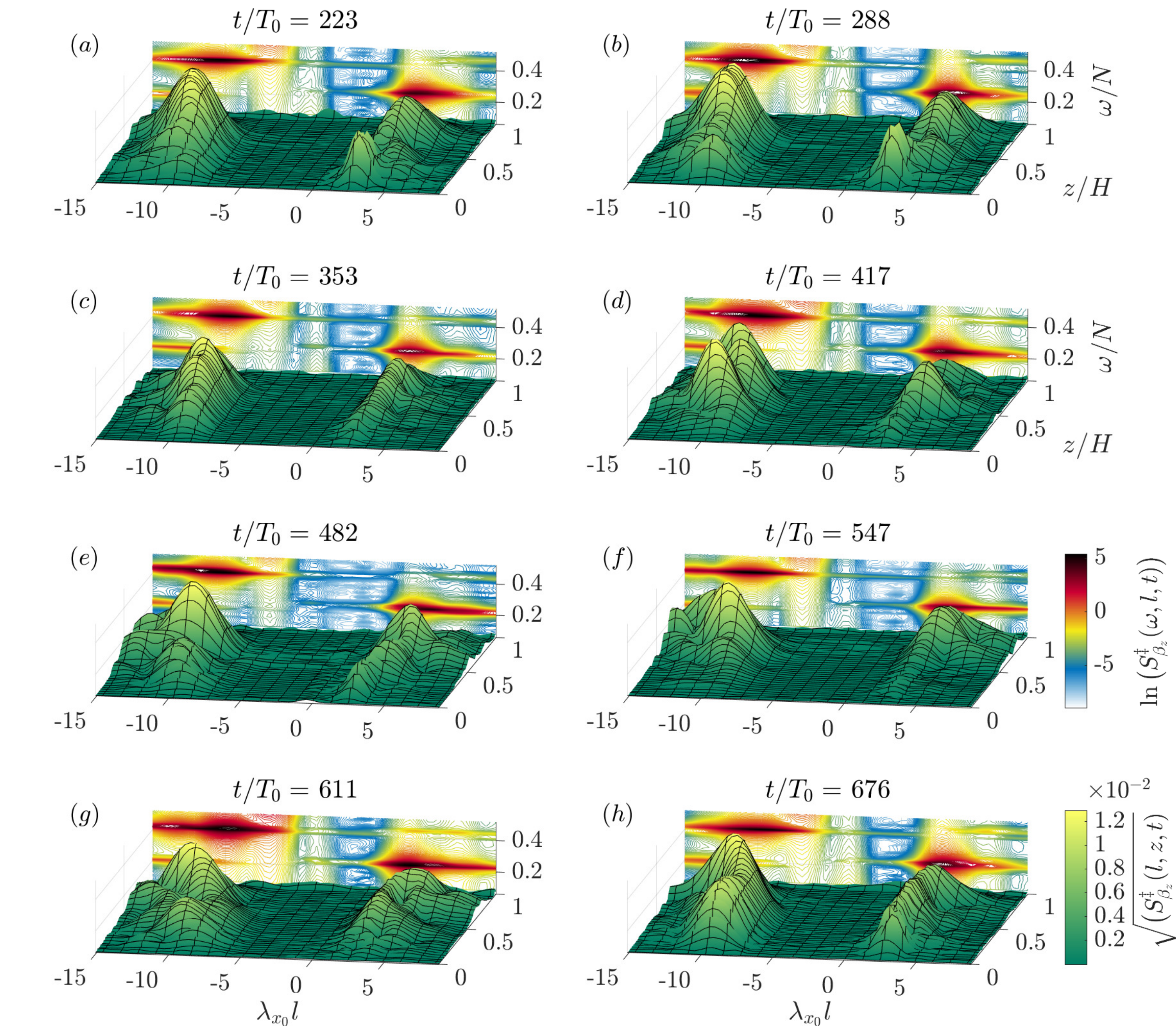}
	\caption{Eight surface plots for eight moments in time of the experiment presented in Figure \ref{fig: time_snaps}, showing the distribution of $l_1$ and $l_2$ over the non-dimensional height in the domain ($z/H$) calculated by (\ref{eqn: surface plot}). Here the $\beta_z$ image sequence is first temporally filtered in Fourier space to remove the signal from $\wave_0$, and thus we do not see a peak at $l_0$. Both the surface plot colour and the height of the peaks, show the power spectral density $S_{\beta_{z}^\ddagger}(l,z)$. The background plot then shows $l_1$ and $l_2$ at the same instant in time in the Fourier plane of horizontal wavenumber component and of frequency. This contour plot is defined by (\ref{eqn: contour plot}). The timings of each image are given by the white dashed lines in Figure \ref{fig: spectrograms}.}
	\label{fig: wavenumber spectrograms}
\end{figure}

The spatial dependence of the instability is highlighted further in Figure \ref{fig: wavenumber spectrograms}, where the three-dimensional surface plots shows the horizontal components of the wave vectors ${l_1}$ and ${l_2}$ as a function of height in the domain for 8 different instances in time, given by the white dashed lines on Figure \ref{fig: spectrograms}(a). The surface is defined by 
\begin{equation}
\label{eqn: surface plot}
S_{\beta_{z}^\ddagger}(l,z,t) = \bigg| \int_{-\infty}^{+\infty}\beta_{z}^\ddagger(x',z,t) e^{-i2\pi l x} W(x' - x; X_X) dx' \bigg|^2,
\end{equation}
where $W(x'; X_X)$ is a Hamming window of width $X_X = x/\lambda_{x_0} =$ 9.7, spanning the full width of the domain. Here, $\beta_{z}^\ddagger(x, z, t)$ corresponds to the instantaneous vertical density perturbation gradient that has already been temporally filtered in Fourier space to remove the signal from $\omega_0$. The surface plots show (\ref{eqn: surface plot}) evaluated at each height in the domain to obtain the horizontal component of wavenumber $l_1$ and $l_2$. The contour plots behind show the corresponding frequency-wavenumber spectrogram. This is obtained by the two-dimensional Fourier transform (in $x$ and $t$)
\begin{equation}
\label{eqn: contour plot}
S_{\beta_{z}^\ddagger}(\omega, l, t) = \bigg\langle\bigg| \frac{1}{T_T}\int\int_{-\infty}^{+\infty}\beta_{z}^\ddagger(x',z,t') e^{-i2\pi(lx' + \omega t')} W(x'-x; X_X) W(t'-t; T_T) dx' dt' \bigg|^2\bigg\rangle_{z},
\end{equation}
where the widths of the Hamming windows are given by $T_T/T_0$ = 39 and $X_X = x/\lambda_{x_0} =$ 9.7, and the subscript $z$ on the angle brackets shows that $S_{\beta_{z}^\ddagger}(\omega, l, t)$ is averaged over the height of the domain. The region of spatio-temporal discontinuity shown in physical space by the black dashed rectangle in Figure \ref{fig: DMD messy}(d) is clearly visible in wavenumber space in Figure \ref{fig: wavenumber spectrograms}(e). Examining the peak of the spectral isosurface, $S_{\beta_{z}^\ddagger}(l,z,t)$, corresponding to $l_2$, around mid-height in the domain, there is a shift in both the amplitude and value of $l_2$ where the peak occurs. The presence of this discontinuous region indicates that two wave beams, of slightly different frequency and wavenumber, are destructively interfering with each other. Later, at $t/T_0$ = 547 in (f), the triadic interaction in the lower part of the domain has decayed (as there is only a very low amplitude signal for both $l_1$ and $l_2$), while the interaction occurring in the upper region of the domain is still present. This continuously varying range of wavenumbers explains why the grey region of experimentally obtained characteristic wavenumbers on Figure \ref{fig: loci_curve} does not exactly fit the spatial triadic conditions of the underlying green branch of the loci. 

We speculate that the reason for these modulations -- observed in both real and Fourier space -- is due to the finite-width of the primary wave beam. As a packet of energy in $\wave_1$ or $\wave_2$ exits the underlying primary beam, the energy exchange between the triad is broken. The time taken for both these secondary beams to exit the spatial confines of $\wave_0$ is dependant on the group velocities of the beams, which are functions of their wavenumbers, and the relative orientation of the beams determined by their frequencies. If the secondary beams are unable to extract sufficient energy before propagating out of the primary beam, the triad system will not be able to form a stable equilibrium and another triadic perturbation will grow in another location. Moreover, all the triadic beams are comprised of a broadband wavenumber spectrum due to their finite-width, as indicated in Figure \ref{fig: wavenumber spectrograms}. This introduces a range of group velocities in the secondary beams which will exit the underlying beam at different times, enhancing the unsteady transfer of energy. 

Additionally, the structure of the underlying $\wave_0$ varies across the height of the domain. As $\wave_0$ propagates upwards through the tank, it decays due to viscosity, resulting in a broadening in spatial extent and reduction in amplitude \citep{Fan2020}. These combine to give considerable variation in both real and Fourier space over the height of the domain, where different locations will favour slightly different triadic perturbations. Indeed, as different perturbations grow, the secondary wave beams with very similar frequencies could interact with each other non-linearly via the primary wave beam, generating a slow ‘beating’ effect. This interaction could cause the secondary beams to decay in some locations, while in others it causes a growth in amplitude, amplifying the effects of the modulations. Making the approximation that there is a single discrete set of parameters corresponding to the secondary wave beam for the whole domain is therefore an oversimplification that ignores the spatial variation of the instability. 


From the experimental results presented above, we believe that the unsteady behaviour of the instability is a function of the finite-width $\Lambda_{0}$ of the primary beam $\wave_0$. We therefore seek to understand this interaction in a two-dimensional context. We pursue this through the development of a two-dimensional weakly non-linear model, which we will refer to as $\Mt$. Details of its development are outlined in the following section. The goal here is to dissect the experiments and to isolate the dynamics that are observed experimentally, in order to improve the understanding of the system. A computational fluid dynamics (CFD) code would be an inappropriate choice to achieve this, as little would be learnt about the physical mechanisms governing the behaviour. In $\S$ \ref{subsec: pert expansion} we examine the perturbation expansion used and in $\S$ \ref{sec: model development} we look at the numerical solution to the obtained system of equations.

\section{Weakly non-linear model construction} \label{sec: numerical set-up}
\subsection{Perturbation expansion}
\label{subsec: pert expansion}

Assuming a two-dimensional incompressible continuously stratified Boussinesq fluid in background hydrostatic balance with constant buoyancy frequency $N$, the full Boussinesq non-linear equations of motion (comprised of the momentum, continuity and mass conservation equations) are given by
%
\begin{align}
\label{eqn: momentum_x}
\varrho \frac{D\boldsymbol{u}}{Dt} &= -\grad p - g\varrho\mathbf{\hat{z}} + \mu \grad^2\boldsymbol{u}, \\
\label{eqn: density_advection}
\frac{D\rho}{Dt} &=  -w\frac{d\bar{\rho}}{dz}, \\
\label{eqn: imcompressiblity NL}
\grad \cdot \boldsymbol{u} &= 0,
\end{align}
where $p$ is the dynamic pressure and $\mu$ is the dynamic viscosity. Writing velocity in terms of the scalar stream function $\boldsymbol{u} = \grad \times (\Psi \hat{\boldsymbol{y}})$, this system can be reduced to
\begin{align}
\label{eqn: final_NonLinear_with_rho}
\frac{D^2}{Dt^2} (\grad^2 \Psi) - \frac{D}{Dt}\nu \grad^2(\grad^2\Psi) &+ N^2\frac{\partial^2\Psi}{\partial x^2} = \frac{g}{\varrho_0}\bigg(\frac{\partial}{\partial x}\frac{D}{Dt}  - \frac{D}{Dt}\frac{\partial}{\partial{x}}\bigg)\rho, \\
\label{eqn: final nonlinar 2nd}
\frac{D\rho}{Dt} &= -\frac{\partial \Psi}{\partial x}\frac{d\bar{\rho}}{dz}.
\end{align}
Here, we focus on (\ref{eqn: final_NonLinear_with_rho}), which represents the non-linear momentum balance in terms of stream function and density. We seek the simplest form of the stream function and density that will describe the behaviour of TRI in a finite-width beam. Specifically we define
\begin{equation}
\label{eqn: psi tilde form}
\Psi = \tilde{\Psi}(x, z, t)e^{i(\boldsymbol{k} \cdot \boldsymbol{x} - \omega t)} + \textrm{c.c.},
\end{equation}
\begin{equation}
\label{eqn: rho tilde form}
\rho = \tilde{\rho}(x, z, t)e^{i(\boldsymbol{k} \cdot \boldsymbol{x} - \omega t)} + \textrm{c.c.},
\end{equation}
where $\tilde{\Psi}$ and $\tilde{\rho}$ are the reduced forms of the stream function and density respectively and c.c. represents the complex conjugate. Both $\tilde{\Psi}$ and $\tilde{\rho}$ are given as functions of space and time because, based on the experimental results, the behaviour of TRI in a finite-width beam is spatio-temporally dependant. We note that in the limit of linear waves we have 
%
\begin{equation}
\label{eqn: rho_as_Psi breve}
\tilde{\rho} = -\frac{l}{\omega} \frac{N^2 \varrho_0}{g} \tilde{\Psi},
\end{equation}
a relationship that also holds for non-linear plane-waves. This relationship will prove useful later in our exploration of the weakly non-linear interactions. 

We define the three non-dimensional parameters 
\begin{align}
\label{eqn: epsilon def 2}
\epsilon &=(|\tilde{\Psi}_{00}|\kappa_0^2) N^{-1}, \\[0.3\baselineskip]
\label{eqn: gamma def}
\gamma &= (\kappa_0 \Lambda_{0})^{-1},\\[0.3\baselineskip]
\label{eqn: chi def}
\chi &=(\nu\kappa_0)\boldsymbol{c}_{g_0}^{-1},
\end{align}
where, $|\tilde{\Psi}_{00}|$ is the characteristic magnitude of the stream function associated with the primary beam $\wave_0$ and $\nu = \mu/\rho$ is the kinematic viscosity. Here, $\epsilon$ can be viewed as a non-dimensional measure of the beam amplitude, which characterises the relative importance of the non-linear $\boldsymbol{u} \cdot \grad$ terms in the momentum and conservation of mass equations. The spatial parameter $\gamma$ defines the separation between the dominant wavelength and overall width of the primary beam. Finally, $\chi$ is the inverse of the Reynolds number defined earlier. In oceanographic settings, $\chi$ will be orders of magnitude smaller than $\epsilon$ and $\gamma$ as, at the scales of interest, the role of viscosity in the ocean can be considered negligible. Here, however, as experiments are inherently more viscous than similar motion patterns at oceanic scales, we retain the leading order effects of viscosity. We shall restrict our attention to low amplitude $\epsilon \sim 10^{-1} \ll 1$, broad beam $\gamma \sim 10^{-1} \ll 1$ and low viscosity $\chi \sim 10^{-2} \ll 1$, and introduce re-scaled time and position through the five variables 
\begin{equation}
\label{eqn: small parameters - time} 
\begin{split}
\tau_\gamma = \gamma t, \qquad \xi &= \gamma x, \qquad \zeta = \gamma z, \\
\qquad \tau_\epsilon = \epsilon t,  &\qquad \tau_{\chi} = \chi t. \\
\end{split}
\end{equation}
As we shall see, $\tau_\epsilon$ accounts for the `slow non-linear time' variations to the amplitude, while $\tau_\gamma$ governs the `slow advection time' scale. As, for the experiments, both $\epsilon \ll 1$ and $\gamma \ll 1$, these scaled times account for change over long-time periods. The final time-scale given by $\tau_{\chi}$ represents the viscous decay of the wave beams. The spatial parameters $\xi$ and $\zeta$ account for the gradual spatial variability of the wave beams in the domain. Utilising the dimensionless amplitude $\epsilon$, we re-write the stream function in (\ref{eqn: psi tilde form}) as 
\begin{equation}
\label{eqn: psi tilde form 2}
\Psi = \tilde{\Psi}(x, z, t)e^{i(\boldsymbol{k} \cdot \boldsymbol{x} - \omega t)} + \textrm{c.c.} = \epsilon\breve{\Psi}(\tau_\gamma, \xi, \zeta, \tau_\epsilon, \tau_{\chi}) e^{i(\boldsymbol{k} \cdot \boldsymbol{x} - \omega t)} + \textrm{c.c.},
\end{equation}
%
where $\tilde{\Psi} = \epsilon\breve{\Psi}$. We therefore require $|\breve{\Psi}| \sim $ 1 as $\tilde{\Psi} = \epsilon \breve{\Psi} = (|\tilde{\Psi}_{00}|\kappa_0^2/N) \breve{\Psi}$ and we are interested in small perturbations to the flow. The `fast time' is associated with the phase variations of the waves given by the wave frequency and is captured by the complex exponential wave form $e^{i(\boldsymbol{k} \cdot \boldsymbol{x} - \omega t)} = e^{i\phi}$. As we are interested in the wave triad, we express the stream function as the summation in the same way as \cite{McEwan1977}, by
\begin{equation}
\label{eqn: triadic streamfunction}
\Psi = \sum_{p=0}^{2}{\Psi}_p =  \sum_{p=0}^{2}\tilde{\Psi}_p(x, z, t)e^{i\phi_p} + \textrm{c.c.} = \sum_{p=0}^{2}\epsilon\breve{\Psi}_p(\tau_\gamma, \xi, \zeta, \tau_\epsilon, \tau_{\epsilon\gamma}, \tau_{\chi}) e^{i\phi_p} + \textrm{c.c.},
\end{equation}
where the subscript $p$ indicates a locally plane-wave approximation to $\wave_0$, $\wave_1$ or $\wave_2$. Each wave phase $e^{i\phi_p}$ therefore represents the characteristic frequency and wavenumber contribution to $\wave_p$. A similar set of expressions can be written for $\rho$ as a sum of $\epsilon \breve{\rho}_p e^{i\phi_p}$.

The superposition in (\ref{eqn: triadic streamfunction}) is then substituted into the non-linear equation (\ref{eqn: final_NonLinear_with_rho}) where, due to the separate space and time-scales, the partial derivatives in (\ref{eqn: final_NonLinear_with_rho}) with respect to $(x,z,t)$ become
\begin{subequations}
	\label{eqn: partial deriv psi}
	\begin{align}
	\label{eqn: partial deriv psi a}
	\begin{split}
	\frac{\partial \Psi_p}{\partial x} &= \epsilon \bigg(i l_p + \gamma \frac{\partial}{\partial \xi} \bigg) \breve{\Psi}_p e^{i\phi_p}, 
	\end{split} \\
	\begin{split}
	\frac{\partial \Psi_p}{\partial z} &= \epsilon \bigg(i m_p + \gamma \frac{\partial}{\partial \zeta} \bigg) \breve{\Psi}_p e^{i\phi_p}.
	\end{split} \\
	\begin{split}
	\frac{\partial \Psi_p}{\partial t} &= \epsilon \bigg(-i \omega_p + \gamma \frac{\partial}{\partial \tau_\gamma} + \epsilon \frac{\partial}{\partial \tau_{\epsilon}} + \chi \frac{\partial}{\partial \tau_{\chi}} \bigg) \breve{\Psi}_p e^{i\phi_p},
	\end{split} 
	\end{align}
\end{subequations}
%
This substitution and following manipulations were performed with the aid of Mathematica \citep{Mathematica2021} to ensure reliability of the lengthy algebraic manipulations required.

We note that at first order in $\epsilon$ the right-hand side (RHS) of (\ref{eqn: final_NonLinear_with_rho}) vanishes. Therefore, as the contributions from density on the RHS first appear at $\mathcal{O}(\epsilon^2)$, the linear relationship in (\ref{eqn: rho_as_Psi breve}) is valid up until order $\mathcal{O}(\epsilon^2)$. The resultant expression obtained provides the Boussinesq viscous equations of motion solely as a function of $\Psi$ that can be used to examine both linear and non-linear dynamics between a triadic set of waves simply by collecting around orders of $\epsilon$, $\gamma$ and $\chi$. 

%
%
%

At $\mathcal{O}(\epsilon^0\gamma^0)$ and $\mathcal{O}(\epsilon^0\gamma^1)$ the expression will be zero as these orders correspond to a state of rest. At order $\mathcal{O}(\epsilon^1 \gamma^0)$, we recover the linear wave solution, with the non-linearity in (\ref{eqn: final_NonLinear_with_rho}) vanishing and linear superposition applying such that the three waves propagate independently. Extracting terms with a common factor of $e^{i\phi_p}$, leaves 
%
\begin{equation}
\frac{\omega_p}{N} = \frac{l_p}{\sqrt{(l_p^2 + m_p^2)}},
\end{equation}
which is the linear dispersion relationship for internal plane-waves. Indeed, for a single non-linear plane-wave in the inviscid limit, the RHS of (\ref{eqn: final_NonLinear_with_rho}) vanishes while the left hand side returns the dispersion relationship for non-trivial solutions, regardless of the wave amplitude. 

We then collect around the next order $\mathcal{O}(\epsilon^1 \gamma^1)$. Again, as $\epsilon$ is still at first-order, the non-linear terms in (\ref{eqn: final_NonLinear_with_rho}) cancel. Looking at the $e^{i\phi_p}$ terms, we obtain 
\begin{equation}
\label{eqn: linear advection split}
\gamma \frac{\partial \breve{\Psi}_p}{\partial \tau_\gamma} = \gamma\bigg(\frac{l_p(\omega_p^2 - N^2)}{\omega_p\kappa_p^2}\frac{\partial \breve{\Psi}_p}{\partial \xi} + \frac{\omega_p m_p}{\kappa_p^2}\frac{\partial \breve{\Psi}_p}{\partial \zeta}\bigg),
\end{equation}
which after some re-arranging can be expressed as 
\begin{equation}
\label{eqn: linear advection}
\gamma \frac{\partial \breve{\Psi}_p}{\partial \tau_\gamma} = -\gamma \big(\boldsymbol{c}_{g_p} \cdot \grad_{\boldsymbol{\xi}}\big) \breve{\Psi}_p \rightarrow \frac{\partial \tilde{\Psi}_p}{\partial t} = - \big(\boldsymbol{c}_{g_p} \cdot \grad\big) \tilde{\Psi}_p,
\end{equation}
where $\grad_{\boldsymbol{\xi}} = (\partial /\partial \xi, \partial /\partial \zeta)$ and $\grad = (\partial /\partial x, \partial /\partial z)$. This linear advection equation shows that the stream function of each wave beam in the triad is advected at its respective group velocity. It is already well known that, for small amplitude internal waves, the group velocity $\boldsymbol{c}_g$ is the velocity at which energy is transported \citep[e.g.][]{Sutherland2010}. As energy scales with $\sim \Psi^2$, the fact that (\ref{eqn: linear advection}) shows that $\breve{\Psi}$ is also advected by $\boldsymbol{c}_g$, is not altogether surprising. 


Before examining the non-linear interaction terms at $\mathcal{O}(\epsilon^2)$, the viscous term at $\mathcal{O}(\epsilon\chi)$ needs to be considered in conjugation with the linear advection recovered at $\mathcal{O}(\epsilon^1 \gamma^1)$. While these terms are of lower magnitude than the terms at $\mathcal{O}(\epsilon^2)$, based on the experimental and oceanographic parameters, they govern the viscous decay of individual wave beams irrespective of the non-linear interactions, making it appropriate to consider their role in conjunction with the advection. At $\mathcal{O}(\epsilon\chi)$, looking at terms with a common factor $e^{i\phi_p}$, we obtain 
\begin{equation}
\label{eqn: linear advection nu}
\chi\frac{\partial \breve{\Psi}_p}{\partial \tau_{\chi}} = - \frac{\boldsymbol{c}_{g_0}}{\kappa_0}\frac{\chi}{2}\kappa_p^2\breve{\Psi}_p \rightarrow \frac{\partial \tilde{\Psi}_p}{\partial t} = - \frac{\nu}{2}\kappa_p^2\tilde{\Psi}_p,
\end{equation}
which shows that the viscous decay of each beam scales as $\kappa_p^2$. We choose to combine the evolution on time-scales $\tau_\gamma$ and $\tau_{\chi}$ to obtain, at first order in $\epsilon$, the advection equation 
\begin{equation}
\label{eqn: linear advection split nu}
\frac{\partial \tilde{\Psi}_p}{\partial t} = -\big(\boldsymbol{c}_{g_p} \cdot \grad \big) \tilde{\Psi}_p - \frac{\nu}{2}\kappa_p^2\tilde{\Psi}_p.
\end{equation}
This advection equation in (\ref{eqn: linear advection split nu}) is solved in the 2D advection component of the $\Mt$ model and its numerical implementation is addressed in \mbox{$\S$ \ref{sec: model development}}.


We next consider terms at order $\mathcal{O}(\epsilon^2 \gamma^0)$. Here, non-linearity enters the problem and terms are no longer only associated with $e^{i\phi_p}$, but are also comprised of cross-terms from $\boldsymbol{u} \cdot \grad$ operator in (\ref{eqn: final_NonLinear_with_rho}), which have the form $e^{i(\phi_q + \phi_r)}$, when expressed in Fourier modes (where $p, q, r$ are permutations of 0, 1, 2). We consider a triad of wave beams satisfying resonance conditions (\ref{eqn:  phase resonant_condition}). Using (\ref{eqn: rho_as_Psi breve}) to eliminate $\breve{\rho}$ from (\ref{eqn: final_NonLinear_with_rho}), at $\mathcal{O}(\epsilon^2)$ we recover the same non-linear interactions given in \citet{Bourget2013}, specifically
\begin{subequations} \label{eqn: zero dimensional ODEs 1}
	\begin{align}
	\label{eqn: zero dimensional ODE0s 2}
	\epsilon^2\frac{\partial \breve{\Psi}_0}{\partial \tau_\epsilon} = \epsilon^2 I_0 \breve{\Psi}_1 \breve{\Psi}_2 \rightarrow \frac{\partial \tilde{\Psi}_0}{\partial t} = I_0 \tilde{\Psi}_1 \tilde{\Psi}_2,  \\
	\label{eqn: zero dimensional ODE1s 2}
	\epsilon^2\frac{\partial \breve{\Psi}_1}{\partial \tau_\epsilon} = \epsilon^2 I_1 \breve{\Psi}_0 \breve{\Psi}_2^* \rightarrow \frac{\partial \tilde{\Psi}_1}{\partial t}  = I_1 \tilde{\Psi}_0 \tilde{\Psi}_2^*, \\
	\label{eqn: zero dimensional ODE2s 2}
	\epsilon^2\frac{\partial \breve{\Psi}_2}{\partial \tau_\epsilon} = \epsilon^2 I_2 \breve{\Psi}_0 \breve{\Psi}_1^* \rightarrow \frac{\partial \tilde{\Psi}_2}{\partial t}  = I_2\tilde{\Psi}_0 \tilde{\Psi}_1^*,
	\end{align}
\end{subequations}
where an asterisk indicates the complex conjugate and the interaction term is given as 
\begin{equation} 
\label{eqn: interaction term}
I_p = \frac{l_q m_r - m_q l_r}{2\omega_p \kappa_p^2} \bigg[\omega_p(\kappa_q^2 - \kappa_r^2) + l_p N^2\bigg(\frac{l_q}{\omega_q} - \frac{l_r}{\omega_r}\bigg)\bigg].
\end{equation}

It is possible to extend this expansion to examine the higher order $\mathcal{O}(\epsilon^2 \gamma^1)$, however as this model sufficiently captures enough of the observed experimental behaviour, this further expansion was not required. Above $\mathcal{O}(\epsilon^2 \gamma^1)$ (higher orders in $\epsilon$) the linear approximation in (\ref{eqn: rho_as_Psi breve}) is no longer valid for eliminating $\breve{\rho}$ from (\ref{eqn: final_NonLinear_with_rho}). The evolution of these coupled ODEs in (\ref{eqn: zero dimensional ODEs 1}), are considered on their own and in $\S$ \ref{subsec: nonlinear interactions} and as part of the $\Mt$ model in $\S$ \ref{subsec: non-linear model results}.

\subsection{Numerical implementation}
\label{sec: model development}

This section outlines the development of the $\Mt$ numerical model, built to solve the equations obtained at order $\mathcal{O}(\epsilon^1 \gamma^1)$ and at order $\mathcal{O}(\epsilon^2)$. We start with the numerical scheme used to solve the advection equation (\ref{eqn: linear advection split nu}), obtained at order $\mathcal{O}(\epsilon^1 \gamma^1)$. Specifically, we use a monotonic second-order upwind finite volume scheme to advect the complex stream function. The finite volume method discretizes the governing equations into arbitrary control volumes around each node and the advective fluxes are then evaluated across the upwind faces of each control volume. For internal waves, their advection velocity is determined by their relative group velocity. As $\boldsymbol{c}_{g_p}$ in (\ref{eqn: linear advection split nu}) is specific to each wave beam, $p$, the number of numerical domains corresponds to the number of wave beams being considered, as, up to $\mathcal{O}(\epsilon^2)$, we only need to consider $\tilde{\Psi}$ and need not explicitly consider $\tilde{\rho}$. There is no limit to the number of superposed domains that can be evolved simultaneously, provided there are suitable interaction terms that couple them. For the results in this paper, we only consider three domains, allowing us to model a three-beam system. Limiting the number of domains to three means we fix the Fourier components of the triad. This turns out to be sufficient to capture the amplitude modulations observed in our experiments, though it excludes frequency and wavenumber fluctuations seen experimentally in the secondary wave beams. 

\begin{figure}
	\centering
	{\includegraphics[width= 0.87\textwidth]{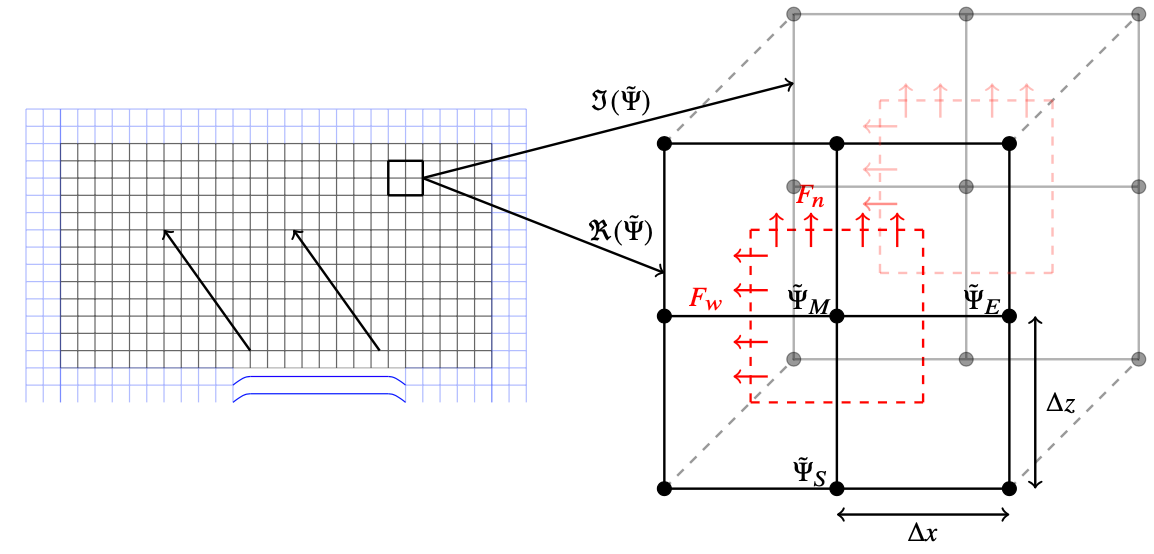}}
	\caption{Sketch outlining the numerical advection domain $\domain_0$. The bottom boundary forcing of the wavemaker is given by the sinusoid envelope in (\ref{eqn: wavebeam envelope}). The two outer grid layers required for the second-order scheme are shown in blue. In this domain $\boldsymbol{c}_{g_0}$ is directed to the north-west, meaning the advection of the complex $\tilde{\Psi}_0$ is to the north-west. For advection, $\tilde{\Psi}_0$  is spilt into its real and imaginary parts and converted back into a complex value after advection is complete. The right panel then show the close up of the real flux (front domain) and imaginary flux (behind domain) over a control volume.}	
	\label{fig: numerical domain}
\end{figure}

We define each domain by $\domain_p$, as we are advecting the reduced stream function $\tilde{\Psi}_p$ in (\ref{eqn: linear advection split nu}). Being a second order scheme, the volume flux, $F$, is calculated using the two upstream values of the reduced stream function. As $\tilde{\Psi}_p$ is complex, the imaginary and real parts are advected separately and combined back into a complex value after being advected. Figure \ref{fig: numerical domain} shows a schematic of the domain $\domain_0$. To avoid confusion with indexing, compass co-ordinates are used for the nodes and the subscript $p=0$ is dropped from $\tilde{\Psi}_0$ in the equations below for clarity. Specifically, for domain $\domain_0$, $F_{w} = c_{g_{0_x}} \tilde{\Psi}_{w}$ and $F_{n} = c_{g_{0_z}} \tilde{\Psi}_{n}$, where
\begin{subequations} \label{eqn: second order upwind}
	\begin{align}
	\tilde{\Psi}_{w} =  \tilde{\Psi}_{M} + \frac{\Upsilon_W}{2}\big(\tilde{\Psi}_{M} - \tilde{\Psi}_{E}\big)	\quad \quad &\textrm{for} \quad 	c_{g_{0_x}} < 0, \leftarrow \\
	\tilde{\Psi}_{n} =  \tilde{\Psi}_{M} +  \frac{\Upsilon_N}{2}\big(\tilde{\Psi}_{M} - \tilde{\Psi}_{S}\big) \quad \quad  &\textrm{for} \quad 	c_{g_{0_z}} > 0, \uparrow 
	\end{align}
\end{subequations}
%
the node indexing is shown in Figure \ref{fig: numerical domain} and $\Upsilon$ is the flux limiter which prevents the generation of oscillations inherent in second-order schemes \citep{Roe1986}. The flux limiter used is the `min-mod' function which takes the ratio of the downstream to upstream gradient. The flux limiter is calculated for both the real and imaginary parts and the more restrictive value (i.e. the one closest to enforcing a first order scheme) is used for both components. Accounting for viscous attenuation, the value of the stream function at the central node $\tilde{\Psi}_{M}$ is then updated by the differences in the fluxes over the control volume using  
\begin{equation} 
\label{eqn: timestepping for psi} 
\tilde{\Psi}_{M}^{\dagger} =  \tilde{\Psi}_M^{i} + \Delta t \bigg(- \grad \cdot \boldsymbol{c}_{g_0} \tilde{\Psi}_M^i  - \frac{\nu}{2}\kappa_0^2\tilde{\Psi}_M^i\bigg) = \tilde{\Psi}_{M}^{i} + \Delta t \Bigg(\frac{F_{w}^i - F_{e}^i}{\Delta x} + \frac{F_{s}^i - F_{n}^i}{\Delta z} - \frac{\nu}{2}\kappa_0^2\tilde{\Psi}_M^i \Bigg),
\end{equation}
where the superscripts $i$ and $\dagger$ represent the value of $\tilde{\Psi}$ before and after advection respectively and $\Delta t$ is the time-step over which code is advanced. For the scheme to remain numerically stable, we limit the time step, $\Delta t$, to satisfy the CFL condition. 

Each domain $\domain_p$ has 242 x 82 cells in the horizontal and vertical respectively, matching the aspect ratio of the experimental visualisation window. The non-dimensional grid spacing $\Delta x / \lambda_{x_0} = \Delta z / \lambda_{x_0} =$ 0.04, is much finer than the smallest wavelength considered. At every time step, random complex background noise of magnitude $10^{-7} \delta e^{i(2\pi \vartheta)}$ is added to each cell in every domain. Both $\delta$ and $\vartheta$ are independent random numbers spanning the range $[0,\hspace{1mm} 1]$, giving a mean magnitude of the order $5 \times 10^{-8}$ with a uniformly distributed phase angle. This background noise is randomly selected at every time step. Without the addition of a perturbation, the instability would not be triggered. Each domain in the system hosts a boundary condition tailored for the wave beam it contains. For the boundaries corresponding to outgoing waves, standard non-reflecting boundary conditions are used. For the incoming boundaries, small amplitude complex noise, of the same structure as the background noise is advected into the domain.


The only domain containing a different inflow boundary is $\domain_0$, where it is necessary to impose bottom boundary forcing to simulate the wavemaker. As the model considers the advection of the complex valued reduced stream function $\tilde{\Psi}$, the boundary condition is given by the envelope of the experimental forcing, which is obtained by removing the fast time forcing, $e^{i\phi_0}$, from (\ref{eqn: wavebeam envelope}) and (\ref{eqn: constant amp forcing}). This boundary condition for $\domain_0$ is shown in Figure \ref{fig: numerical domain}. We implement a prescribed displacement condition by computing the corresponding complex valued stream function amplitude $\tilde{\Psi}_\textrm{in}$, as opposed to $\eta_0$, where  $\tilde{\Psi}_\textrm{in}$ = $|\tilde{\Psi}_\textrm{in}|e^{i(2\pi \varsigma)}$. Here, $\varsigma$ is a constant value within the range $[0,\hspace{1mm} 1]$. This ensures that the phase angle of the slowly evolving complex amplitude $\tilde{\Psi}_0$ is fixed at the boundary, in the same way as the experimental wavemaker. At each time step, after each domain is advected at its respective group velocity, the non-linear interactions (\ref{eqn: zero dimensional ODEs 1}) are calculated. 

The non-linear interactions are included by converting (\ref{eqn: zero dimensional ODEs 1}) into the numerical format 
%
%
\begin{subequations} \label{eqn: non linear interactions model}
	\begin{align}
	\label{eqn: non linear interactions model 1}
	\tilde{\Psi}_{0}^{i+1} &= \tilde{\Psi}_{0}^{\dagger} + \Delta tI_0 \tilde{\Psi}_{1}^{\dagger} \tilde{\Psi}_{2}^{\dagger},  \\
	\label{eqn: non linear interactions model 2}
	\tilde{\Psi}_{1}^{i+1} &= \tilde{\Psi}_{1}^{\dagger} + \Delta tI_1 \tilde{\Psi}_{0}^{\dagger} \tilde{\Psi}_{2}^{\dagger*},  \\
	\label{eqn: non linear interactions model 3}
	\tilde{\Psi}_{2}^{i+1} &= \tilde{\Psi}_{2}^{\dagger} + \Delta tI_2 \tilde{\Psi}_{0}^{\dagger} \tilde{\Psi}_{1}^{\dagger*},
	\end{align}
\end{subequations}
where $\tilde{\Psi}_p$ corresponds to every cell in domain $\domain_p$ and $\dagger$ and  $i+1$ respectively represent the values of $\tilde{\Psi}_p$ in each domain after advection (\ref{eqn: timestepping for psi}) and at time $t + \Delta t$ after the non-linear interactions are calculated. As the interaction co-efficient is applied to the whole domain, there is no need to distinguish here the different cells using compass indexing. In the following section $\S$ \ref{subsec: nonlinear interactions} we first consider the non-linear interactions on their own, before examining the results from the $\Mt$ model in $\S$ \ref{subsec: non-linear model results}.


\section{Weakly non-linear behaviour} 
\label{sec: nonlinear interactions} 

\subsection{Zero-dimensional model results} 
\label{subsec: nonlinear interactions} 

We note that the coupled non-linear equations in(\ref{eqn: zero dimensional ODEs 1}) can be recovered in zero-dimensional space by considering the form of the stream function $\tilde{\Psi}_p(t) e^{i\phi_p}$ in (\ref{eqn: final_NonLinear_with_rho}), where the slowly evolving reduced stream function is considered solely as a function of time. Indeed, this was first proposed by \cite{McEwan1977} and further developed by \cite{Koudella2006} and \cite{Bourget2013}, who obtain these coupled ODEs that govern the development of the triad. 

While this theory considers the `slow-time' development of the amplitude of the beams, as noted by \citet{Sutherland2013}, it is still based upon the assumption that the waves are monochromatic in space and time. In the case of a finite-width beam becoming unstable to TRI, the secondary wave beams have a finite time with which to interact with the underlying primary beam. This limitation was addressed by \citet{Bourget2014}, who adapted the equations in (\ref{eqn: zero dimensional ODEs 1}) to examine the energy flux across a finite region of the primary wave beam. Using an energy balance, they define a two-dimensional control area of width $W$ and length $L$ over which the resonant beams can interact with the primary. Accounting for the energy flux through this control area via non-linear interactions, viscous attenuation and incoming and outgoing energy flux of the primary beam, the ODEs in (\ref{eqn: zero dimensional ODEs 1}) become 
\begin{subequations} \label{eqn: zero dimensional ODE - finite width}
	\begin{align}
	\label{eqn: zero dimensional ODE1 - finite width}
	\frac{d\tilde{\Psi}_0}{d t} &= I_0 \tilde{\Psi}_1 \tilde{\Psi}_2 - \nu\bigg(\frac{\kappa_0^2}{2} \bigg)\tilde{\Psi}_0 + T, \\
	\label{eqn: zero dimensional ODE2 - finite width}
	\frac{d\tilde{\Psi}_1}{d t}  &= I_1 \tilde{\Psi}_0 \tilde{\Psi}_2^* - \bigg(\frac{\nu \kappa_1^2}{2} + \frac{|\boldsymbol{c}_{g_1} \cdot \boldsymbol{e}_{k_0}|}{2W}\bigg)\tilde{\Psi}_1, \\
	\label{eqn: zero dimensional ODE3 - finite width}
	\frac{d\tilde{\Psi}_2}{d t}  &= I_2 \tilde{\Psi}_0 \tilde{\Psi}_1^* - \bigg(\frac{\nu \kappa_2^2}{2} + \frac{|\boldsymbol{c}_{g_2} \cdot \boldsymbol{e}_{k_0}|}{2W}\bigg)\tilde{\Psi}_2,
	\end{align}
\end{subequations}
where $I_p$ is given in (\ref{eqn: interaction term}),  $\boldsymbol{e}_{k_0}$ is a unit vector in the direction of $\boldsymbol{k}_0$ and the forcing term $T = |\boldsymbol{c}_{g_0}|(\tilde{\Psi}_{\textrm{in}}^*\tilde{\Psi}_\textrm{in} - \tilde{\Psi}_0^*\tilde{\Psi}_0)/(2L\tilde{\Psi}_0^*)$ in (\ref{eqn: zero dimensional ODE1 - finite width}), represents the energy flux for the primary wave beam through the control area with an incoming amplitude of $\tilde{\Psi}_{\textrm{in}}$. We convert this input amplitude to the non-dimensional measure $\epsilon_{\textrm{in}} = \kappa_0^2|\tilde{\Psi}_{\textrm{in}}|/N$. We recognise that $\epsilon_{\textrm{in}}$ is comparable to the experimental input amplitude $\epsilon_r = \kappa_0^2\langle|\tilde{\Psi}_0|\rangle_r/N $, where $\langle|\tilde{\Psi}_0|\rangle_r$ is magnitude of the reduced stream function averaged over the black domain in Figure \ref{fig: DMD TRI breakdown}(b). The terms on the end of (\ref{eqn: zero dimensional ODE2 - finite width}) and (\ref{eqn: zero dimensional ODE3 - finite width}) represent the viscous decay within, and flux of energy out of, the control area. For the remainder of this paper, we will refer to the above set of spatially zero-dimensional ODEs in (\ref{eqn: zero dimensional ODE - finite width}), which we use to describe the energy exchange in TRI in the context of a finite-width beam, as the zero-dimensional model $\Mz$.

The $\Mz$ model is numerically integrated to examine its prediction for the development of the triad. To match the experimental set-up, the width $W$ and length $L$ of the interaction region are set to $\Lambda_{0}$ (defined in (\ref{eqn: delta width})) and $2\Lambda_{0}$ respectively (see \cite{Bourget2014} for details). The parameters for $\wave_0$ are also kept consistent with the experimental ones. In the limit of $\Mz$, $\wave_0$ reduces to 
\begin{equation}
\label{eqn: wave0 input} 
\wave_0 = \{ \omega_0/N = 0.62, \hspace{2mm} \lambda_{x_0}\boldsymbol{k}_0 = (-6.28,  -8.29) \}.
\end{equation}

We are curious to see how varying the wavenumbers and frequencies of the secondary wave beams impact the evolution of the instability described by the $\mathcal{M}_{0\textrm{D}}$. This is achieved by defining a resonant triad as $\pair_\Phi = \{\wave_0, \wave_{1_\Phi}, \wave_{2_\Phi} \}$, where the subscript $\Phi$ corresponds to a specific triadic configuration, obtained by changing the characteristic frequencies and wavenumbers of the secondary wave beams. We require that all the triadic configurations satisfy both the resonant condition (\ref{eqn: resonant_condition}) and dispersion (\ref{eqn: dispersion}), meaning all of the configurations lie exactly on the green curve in Figure \ref{fig: loci_curve}.  For the $\Mz$ model we consider configurations $\pair_a$ and $\pair_d$, marked by the blue circle and star on Figure \ref{fig: loci_curve}, respectively, with parameters shown in Table \ref{table: input paramets for secondary waves 2}. 

\begin{table}
	\centering
	\begin{tabular}{c | c | c | c | c | c | c}
		Triad configuration $\pair_{\Phi}$ & {\hspace{2mm} $\omega_1/N$ \hspace{2mm} } &\hspace{2mm}  {$\omega_2/N$} \hspace{2mm}  & {\hspace{2mm}  $\lambda_{x_0} l_1$\hspace{2mm}  } & {\hspace{2mm}  $\lambda_{x_0}  l_2$ \hspace{2mm} } & {\hspace{2mm} $\lambda_{x_0}  m_1$ \hspace{2mm} } & {\hspace{2mm} $\lambda_{x_0} m_2$ \hspace{2mm} } \\ 
		$\pair_a$ & 0.198 & 0.419 & 2.136 & $-$ 8.671 & 10.304 & $-$18.598
		\\   
		$\pair_b$  & 0.206 &0.411 &  2.388 & $-$8.922 & 11.561 & $-$19.855
		\\   
		$\pair_c$  & 0.222 & 0.395 & 3.267 & $-$9.802 &  14.577 & $-$22.871
		\\ 
		$\pair_d$  & 0.227 & 0.390 & 3.644 & $-$10.179 & 15.708 & $-$24.127
		\\ 
		$\pair_e$  & 0.231 & 0.386 & 4.021 & $-$10.556 & 16.965 & $-$25.258
		\\
		$\pair_f$  & 0.239 & 0.378 & 4.775 & $-$11.310 & 19.227 & $-$27.520
		\\
	\end{tabular}
	\caption{The input parameters of $\wave_{1_\Phi}$ and $\wave_{2_\Phi}$ for each $\pair_{\Phi} = \{\wave_0, \wave_{1_\Phi}, \wave_{2_\Phi}\}$,  used in the $\Mz$ and $\Mt$ model. The wave vector locations of each secondary wave beam pair can be seen by the blue marks on Figure \ref{fig: loci_curve}.}
	\label{table: input paramets for secondary waves 2}
\end{table}

\begin{figure}
	{\includegraphics[width= 1\textwidth]{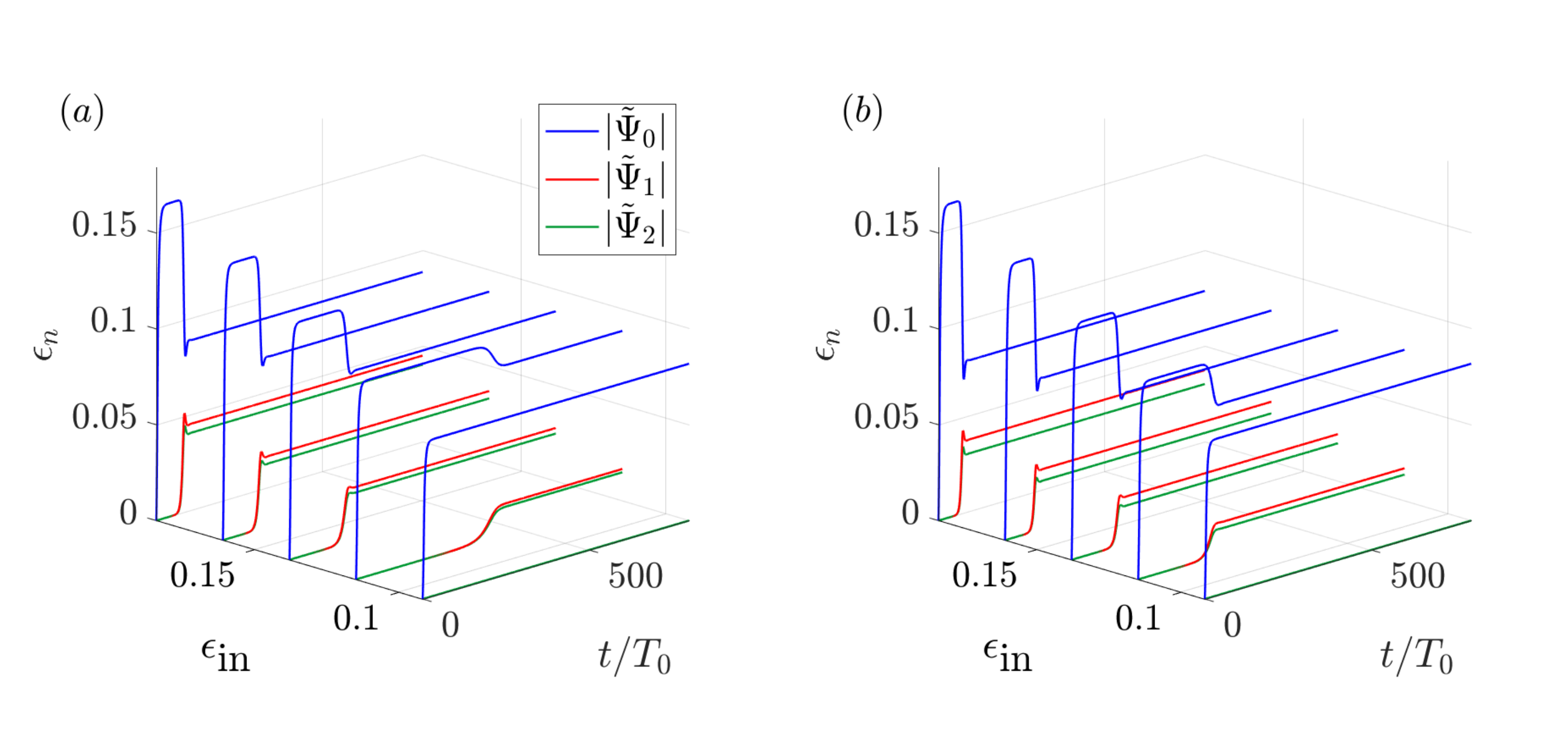}}
	\caption{Results of the $\mathcal{M}_{0\textrm{D}}$ model by \citep{Bourget2014} given in (\ref{eqn: zero dimensional ODE - finite width}) across a range of forcing amplitudes $0.092 \leq \epsilon_{\textrm{in}} \leq 0.184$. (a) corresponds to the triad configuration $\pair_a$, while (b) corresponds to $\pair_d$. The parameters for these configurations are given in Table \ref{table: input paramets for secondary waves 2}.}
	\label{fig: 0D french}
\end{figure}

The results of the numerical integration for the $\Mz$ model are given in Figure \ref{fig: 0D french}(a) and (b) for $\pair_a$ and $\pair_d$, respectively, across a range of five non-dimensional input forcing amplitudes for the primary beam $0.092 \leq \epsilon_{\textrm{in}} \leq $0.184. The resultant amplitudes of the triadic beams are given in terms of $\epsilon_n = \kappa_0^2|\tilde{\Psi}_p|/N$. In order to draw the most meaningful comparison between the spatially 0D model and 2D experiments, we average the experimental results over the whole visualisation window $\langle \rangle_w$ in order to de-correlate the signal with the position of a beam in space. As the $\Mz$ model is not a function of space, no spatial averaging is required for $\epsilon_n$. While $\epsilon_n$ and $\epsilon_w$ are therefore not quantitatively comparable, they are qualitatively. For details on the different measures of amplitude, see Table \ref{table: amplitude}. 





%


We first consider the general behaviour of the $\Mz$ model, consistent between both Figure \ref{fig: 0D french}(a) and (b). For the lowest forcing amplitude of $\epsilon_{\textrm{in}}$ = 0.092, no instability occurs, so $|\tilde{\Psi}_1|$ (red line) and $|\tilde{\Psi}_2|$ (green line) remain zero while $|\tilde{\Psi}_0|$ (blue line) increases with the initial growth of $|\tilde{\Psi}_\textrm{in}|$ and then remains at a constant amplitude. There then exists a critical forcing amplitude, above which, TRI is observed. For $\pair_a$ (Figure \ref{fig: 0D french}(a)) this occurs when $\epsilon_{\textrm{in}} \geq$ 0.111, while for $\pair_d$ (Figure \ref{fig: 0D french}(b)), instability occurs when $\epsilon_{\textrm{in}} \geq$ 0.097. While both of these limits are lower than the experimental limit of \mbox{$\epsilon_r \geq$ 0.166}, the $\Mz$ still captures the fact that there exists an amplitude threshold that must be surpassed before a finite-width beam exhibits TRI. We next note, that for all solutions where TRI is observed, the amplitude of $|\tilde{\Psi}_0|$ always decays to the same asymptotic value, while the amplitudes of $|\tilde{\Psi}_1|$ and $|\tilde{\Psi}_2|$ asymptote at higher values as $\epsilon_{\textrm{in}}$ is increased. Upon investigation, the asymptotic amplitude of $|\tilde{\Psi}_0|$ after the initial decay is equal to the amplitude threshold for instability.  Interestingly, for the larger forcing amplitudes, the approach to the equilbiurm state takes the form of an under-damped non-linear oscillator, shown by a small oscillation after the initial onset of the instability. 

The difference between how the triad configurations $\pair_a$ and $\pair_d$ affect the evolution of the instability is subtle. On inspection, Figure \ref{fig: 0D french}(b) shows a quicker growth of the secondary beams for the same input amplitude as $\pair_a$, a result that agrees with $\pair_d$ having the lower amplitude threshold for instability. Strangely, despite this quicker growth, the resultant amplitudes of the secondary wave beams are smaller than in Figure \ref{fig: 0D french}(a). This is unintuitive, as one would expect that a larger decay of the primary beam would result in larger amplitudes of the secondary beams. As the viscous decay along a beam scales with $\kappa^3$, this lower value of the secondary beams might be due to greater viscous dissipation in $\pair_d$, due to $\kappa_d > \kappa_a$. 

\begin{table}
	\centering
	\begin{tabular}{c | c | c | c | c | c | c} 
		\multicolumn{1} {p{1.6cm}|} {\centering non-dimensional \\ amplitude }  & \multicolumn{1} {|p{2cm}|} {\centering Experimental (E) Modelling ($\Mt$ or $\Mz$)} & \multicolumn{1} {|p{1.8cm}|} {\centering dimensional \\ amplitude } & {\hspace{2mm}  units \hspace{2mm} } &  \multicolumn{1} {p{6cm}} {\centering Description }  \\ 
		$\eta_0 \lambda_{x_0}$ & E & $\eta_0$ & mm & \multicolumn{3} {p{6cm}} {\centering Imposed half peak to peak displacement \\ from the wavemaker  } 
				\\   
		$\epsilon_r$  & E & $\langle|\tilde{\Psi}_0|\rangle_r$ & mm s$^{-2}$ & \multicolumn{3} {p{6cm}} {\centering Spatially averaged measure of $\wave_0$ over \\ region shown in Figure \ref{fig: DMD TRI breakdown}(b)} 
		\\   
		$\epsilon_w$  & E $\&$ $\Mt$ & $\langle|\tilde{\Psi}_p|\rangle_w$  &  mm s$^{-2}$ & \multicolumn{3} {p{6cm}} {\centering Spatially averaged measure of all wave beam fields over their respective domains } 
		\\ 
		$\epsilon$  & $\Mt$ & $|\tilde{\Psi}_{00}|$  & mm s$^{-2}$ & \multicolumn{3} {p{6cm}} {\centering Characteristic measure of primary wave amplitude for perturbation expansion } 
		\\ 
		$\epsilon_{\textrm{in}}$  & $\Mt$ $\&$ $\Mz$ & $|\tilde{\Psi}_{\textrm{in}}|$ & mm s$^{-2}$ & \multicolumn{3} {p{6cm}} {\centering Imposed amplitude input for $\domain_0$ ($\Mt$) and $\wave_0$ ($\Mz$)} 
		\\
		$\epsilon_n$  & $\Mz$ & $|\tilde{\Psi}_p|$ & mm s$^{-2}$ & \multicolumn{3} {p{6cm}} {\centering Measure of amplitude for all wave fields \\ from $\Mz$} 
		\\
	\end{tabular}
	\caption{Details of the different non-dimensional measures of amplitude for clarity.}
	\label{table: amplitude}
\end{table}

Comparing Figure \ref{fig: 0D french} with the experimental results in Figure \ref{fig: amplitude modulation}, we see remarkably different behaviour in the evolution of the instability. For the $\Mz$ model results, larger forcing engenders a significant decay in amplitude of the primary wave beam. This large decay of the primary beam is not observed experimentally, where the amplitude of the primary beam oscillates around a mean value comparable with that set by the forcing from the wavemaker. The most obvious difference between the two then arises in their description of the long-term development. The $\Mz$ model results predict that after the initial instability, the energy exchange between the triad saturates to a steady equilibrium, set by the non-linear interaction term $I$. This is clearly not the case for the experimental results. The slow synchronous amplitude modulations seen experimentally reveal a continuous fluctuation in the energy exchange between the primary and the two secondary beams. 

As the $\Mz$ model does not consider the triadic interaction as a function of space it is unable to describe the modulations witnessed experimentally. We next consider the results of the $\Mt$ model to see how it describes the instability when considered in a two-dimensional domain. 

\subsection{Two-dimensional model results} 
\label{subsec: non-linear model results}

We now consider the results from the $\Mt$ model, where in each simulation we provide three domains -- one for each triadic beam -- with input parameters chosen to satisfy the triadic condition (\ref{eqn:  phase resonant_condition}) and dispersion (\ref{eqn: dispersion}). Six triad configurations $\pair_\Phi = \{\wave_0, \wave_{1_\Phi}, \wave_{2_\Phi} \}$ are considered, with wavenumber vectors distributed across the solid green loci branch in Figure \ref{fig: loci_curve} and parameters given in Table \ref{table: input paramets for secondary waves 2}. The parameters for $\wave_0$ (common to all six triads), match the experiments and are given in (\ref{eqn: wave0 input}).
%



\begin{figure}
	\centering
	\includegraphics[width=0.95\textwidth]{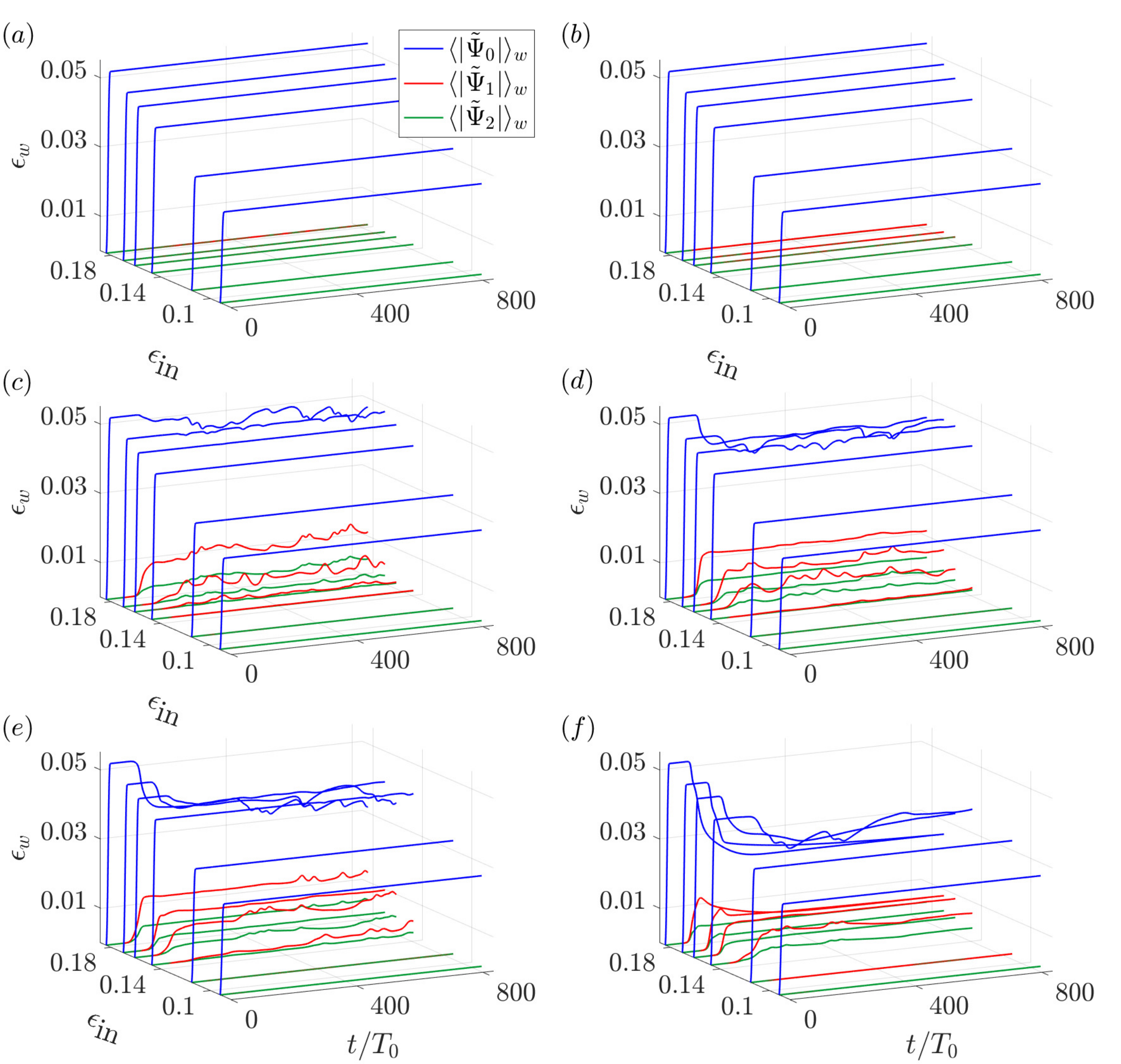}
	\caption{Amplitude plots generated using the two-dimensional weakly non-linear $\Mt$ model. Each sub-plot shows a range forcing amplitudes for $\wave_0$ between \mbox{$0.092 \leq \epsilon_\textrm{in} \leq 0.184$}. The difference between each sub-plot is the parameters for the secondary wave beams in each triad, given in Table \ref{table: input paramets for secondary waves 2}. (a) $\pair_a$ (b) $\pair_b$ (c) $\pair_c$ (d) $\pair_d$ (e) $\pair_e$ (f) $\pair_f$. The run time is $t_{\textrm{end}}/T_0 = 816$. The resultant non-dimensional amplitudes are calculated over the whole visualisation window $\epsilon_w = \kappa_0^2\langle |\tilde{\Psi}|\rangle_w/N$.}
	\label{fig: amp35_outerbranch}
\end{figure}

Figure \ref{fig: amp35_outerbranch} shows the results of 36 simulations, where sub-plots (a) to (f) correspond to triad configurations $\pair_{a}$ to $\pair_{f}$ respectively, each shown with six input amplitudes, $\epsilon_{\textrm{in}}$. Again, $\epsilon_{\textrm{in}}$ is comparable to the experimental values $\epsilon_r$. Looking at all of the plots in Figure \ref{fig: amp35_outerbranch}, it is relatively easy to categorise three different behavioural evolutions of the triad simulations. The first behavioural evolution, observed for all the simulations using triad configurations $\pair_a$ and $\pair_b$ (shown by the blue circle and cross on Figure \ref{fig: loci_curve}), is when no growth of the secondary wave beams occur and the system remains as a single stable primary beam. The reason for this is twofold: both triad configurations satisfy (or nearly satisfy in the case of $\pair_b$) $\kappa_1 / \kappa_0 \leq 1$ and both correspond to the smallest values of $\omega_1$ considered. Indeed, for input amplitudes that caused the other triadic configurations to become unstable, we found that no pairs with $\kappa_1 / \kappa_0 < 1$ generated TRI for the $\Mt$ model. This observation is further reinforced by our experiments: the grey shaded region on Figure \ref{fig: loci_curve} shows that all observed triadic combinations of wave vectors satisfy $\kappa_1 / \kappa_0 > 1$. When $\kappa_1 / \kappa_0 < 1$, we have the condition $\kappa_1 < \kappa_0 < \kappa_2$. Thus energy transfers to both the larger length scale ($\kappa_1$) and the smaller ($\kappa_2$). \cite{Bourget2014} show that the linear growth rate from the $\Mz$ model is larger for this triad configuration when $\Lambda_{0} < 7\lambda_{0}$. However, in our experiments where $\Lambda_{0} \approx 3\lambda_{0}$, it is still a triad configuration with $\kappa_1 / \kappa_0 > 1$ (and therefore $\kappa_0 < \kappa_1, \kappa_2$, located on the outer solid green branch of Figure \ref{fig: loci_curve}), that is selected. The effect of having a smaller value of $\omega_1$  is discussed further below.  

The second behavioural evolution seen in Figure \ref{fig: amp35_outerbranch}, is when TRI occurs and the amplitudes of all beams undergo coupled modulations, similar to those seen experimentally in Figure \ref{fig: amplitude modulation}. This behaviour is observed for many of the simulations using triad configurations $\pair_c$ to $\pair_f$. For these configurations, we see that an amplitude threshold must be surpassed before instability can occur. We note a very close agreement in the amplitude threshold required for instability between $\Mt$ and the experimental values. For the triad configurations $\pair_c$ to $\pair_f$, instability occurred when $\epsilon_{\textrm{in}} \geq$ 0.161, while experimentally $\epsilon_r \geq$ 0.166 triggered instability. We also notice that for all of these simulations that become unstable, as we increase $\epsilon_{\textrm{in}}$, not only do the secondary wave beam amplitudes increase, their growth also occurs at earlier times. This observation is consistent with the $\Mz$ model. We can measure this initial linear growth using the growth rate $\sigma$, of the form $e^{\sigma t}$, to characterise how quickly the instability develops. This growth rate term will prove useful later. 

\begin{figure}
	\centering
	\begin{tikzpicture}
	\node at (1,1) {\includegraphics[width=1\textwidth]{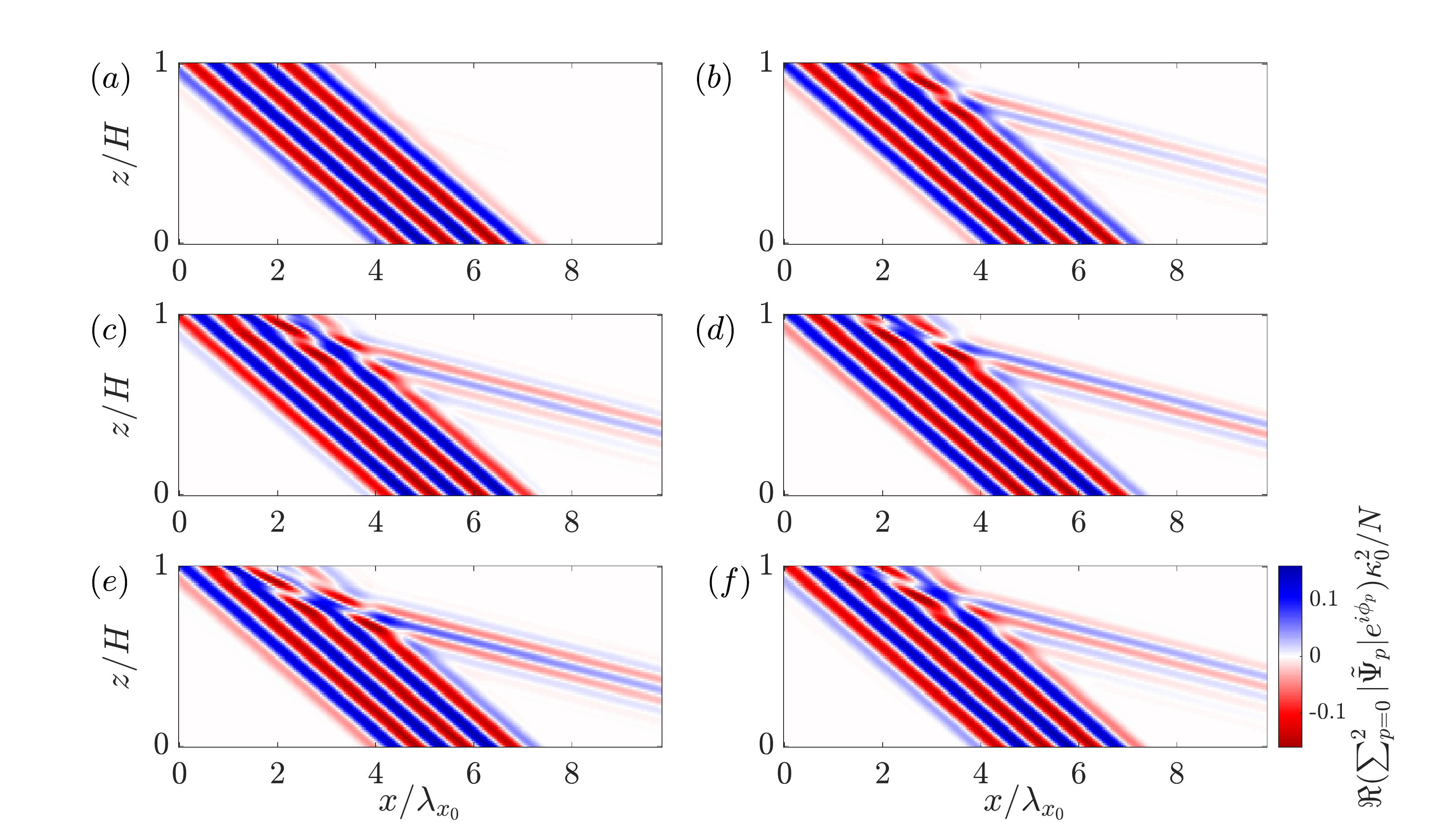}};
	\draw[black, thick, <->]  (1.75, -0.6) -- (3.7, -1.15);
	\node[black] at (2, -1) {\footnotesize $\delta_1$};	
	\end{tikzpicture}
	\caption{Sequence of images at $t/T_0 = 61$ apart, showing the superposition of the three domains in the model, using triad configuration $\pair_d$. Each domain is multiplied by its respective short length and fast time-scales $e^{i\phi_p}$. The forcing amplitude for $\wave_0$ is $\epsilon_\textrm{in}$ = 0.161. The spatially averaged amplitude of each wave beam over time is shown in Figure \ref{fig: amp35_outerbranch}(d). The timing of each image is \mbox{(a) $t/T_0$ = 121,} \mbox{(b) $t/T_0$ = 151,} (c) $t/T_0$ = 272, (d) $t/T_0$ = 333, (e) $t/T_0$ = 424, (f) $t/T_0$ = 484. The black line in (f) marks the length $\delta_1$, which defines the spatial distance over which $\wave_1$ can extract energy from $\wave_0$.}
	\label{fig: time snaps outer branch simple}
\end{figure}

Focusing on a specific simulation with coupled amplitude modulations, we consider triadic configuration $\pair_d$, forced at a non-dimensional input amplitude of $\epsilon_\textrm{in}$ = 0.161 (Figure \ref{fig: amp35_outerbranch}(d)). Figure \ref{fig: time snaps outer branch simple} shows six instantaneous images from this simulation obtained by the superposition of the three domains multiplied by their respective fast time and short length scales $e^{i\phi_p}$. Here we see the initial development of $\wave_1$ occurring at the top of $\wave_0$, where it grows in strength. (We note that calculations preformed in a larger domain demonstrated that the generation region of $\wave_1$ occurs at the same location, showing that it is not an effect from the boundary.) As with the experimental images in Figure \ref{fig: time_snaps}, due to the similar alignment and direction of $\boldsymbol{k}_0$ and $\boldsymbol{k}_2$, $\wave_2$ is not obvious in this visualisation region as it propagates within the confines of $\wave_0$. Over time, $\wave_1$ grows in both amplitude and width before decaying in a quasi-periodic manner. Unlike the experimental results, however, its generation region remains approximately fixed and does not traverse the height of $\wave_0$. By construction, $\omega_1$ and $\omega_2$ also remain fixed.

The final behavioural evolution seen in Figure \ref{fig: amp35_outerbranch}, most obvious for the largest forcing amplitude using $\pair_f$, is when the triadic system reaches a stable equilibrium, closely resembling the steady state results from $\Mz$ shown in Figure \ref{fig: 0D french}. Here, the amplitudes of the triadic beams do not exhibit any modulations and the triad quickly reaches a stable equilibrium, after a large, smooth decay in amplitude of the primary beam.

\begin{figure}
	{\includegraphics[width= 0.9\textwidth]{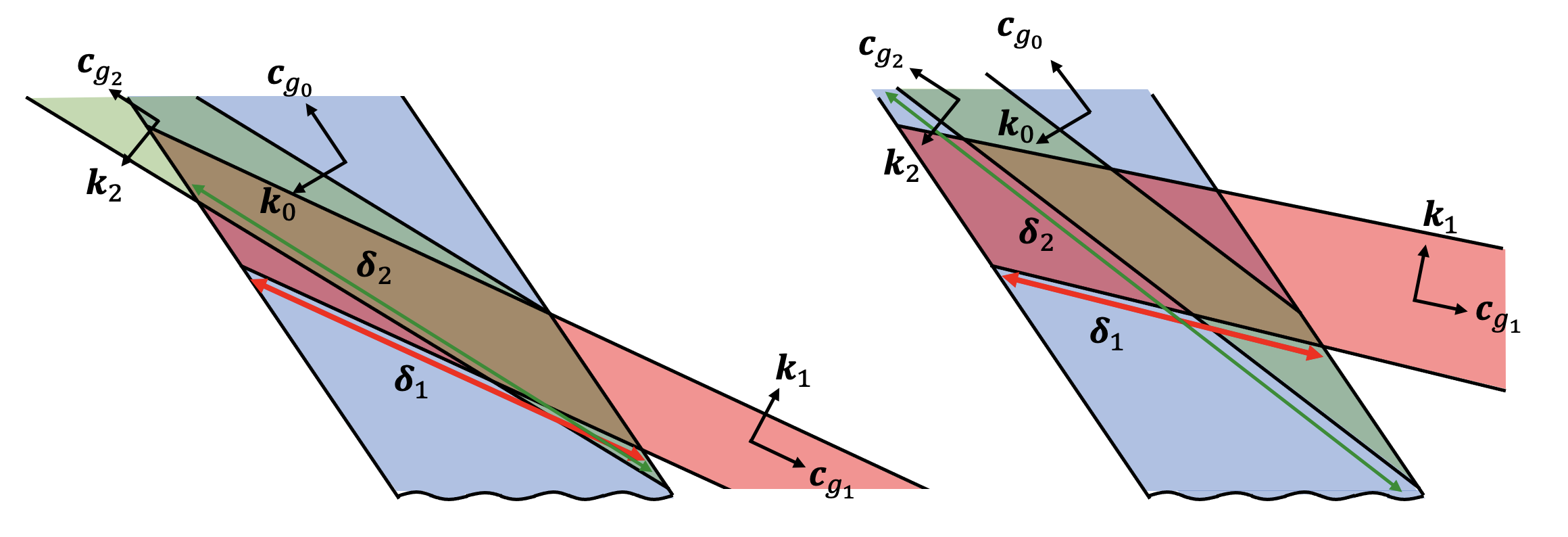}}
	\caption{Schematic showing the effect of different $\omega_1$ and $\omega_2$ combinations on $\delta_1$ (red) and $\delta_2$ (green), the distances over which the secondary wave beams can extract energy with the primary beam before exiting the boundary.}
	\label{fig: wave_config}
\end{figure}

\begin{figure}
	\centering
	\includegraphics[width=0.9\textwidth]{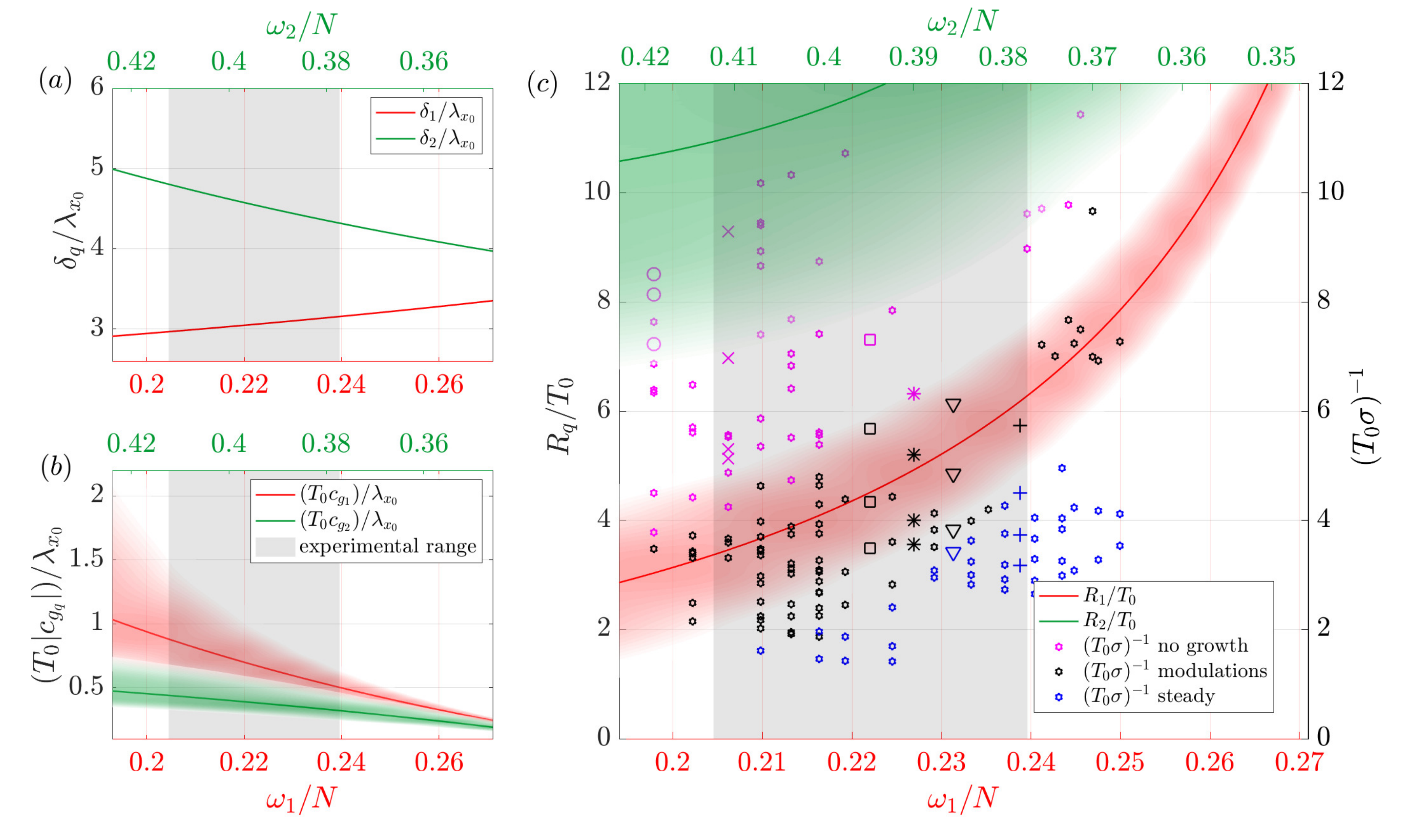}
	\caption{(a)  The non-dimensional interaction length of both $\wave_1$ and $\wave_2$ with $\wave_0$, again as a function of non-dimensional frequency. The grey region in the background highlights the range of experimentally obtained frequencies. (b) The non-dimensional group velocity of both $\wave_1$ (red) and $\wave_2$ (green) as a function of non-dimensional frequency. The red and green shaded regions corresponds to the range of $|\boldsymbol{c}_g|$ possible for a fixed frequency due to a broadband wavenumber spectrum. Details of how these ranges are calculated are provided in the text. (c) The non-dimensional residence time $R_1/T_0$ (red) and $R_2/T_0$ (green) as a function of non-dimensional frequency, along with the non-dimensional inverse linear growth rates $(T_0 \sigma)^{-1}$, for the simulations shown in Figure \ref{fig: amp35_outerbranch} and many others. The style of the growth rate marker indicates the triad configuration tested (given in Figure \ref{fig: loci_curve}), while smaller hexagons are for other simulations not shown in Figure \ref{fig: amp35_outerbranch}). The colour of the marker indicates the behaviour of the simulation, characterised as no growth of secondary waves (magenta), amplitude modulations to the triadic beams (black) or steady equilbiurm of all amplitudes in the triad (blue). The red and green shaded regions associated with the residence time are given from the range of $|\boldsymbol{c}_g|$ shown in (b).}
	\label{fig: interaction time}
\end{figure}

The only parameters being varied across these simulations in Figure \ref{fig: amp35_outerbranch} are the frequencies and wavenumbers of the secondary beams in the triad and the amplitude of the primary beam. This varying behavioural evolution of the triadic beams must, therefore, be due to these changing parameters. Based on the dispersion relationship (\ref{eqn: dispersion}), a greater value of $\omega_1$ results in a $\wave_1$ beam with closer alignment to $\wave_0$. This steeper angle leads to a greater spatial region over which $\wave_1$ can extract energy from $\wave_0$ and consequently an increased distance for $\wave_1$ to grow. This is shown schematically in Figure \ref{fig: wave_config}. We define the lengths of these interaction regions for $\wave_1$ and $\wave_2$ as $\delta_1$ and $\delta_2$, respectively. Figure \ref{fig: interaction time}(a) shows the relationship between $\delta_1$ and $\omega_1$, together with the relationship between $\delta_2$ and $\omega_2$. The grey shading in the background marks the range of frequencies obtained experimentally. 

As $\omega_1$ increases, not only does $\delta_1$ increase, $|\boldsymbol{c}_{g_1}|$ also decreases, as shown in Figure \ref{fig: interaction time}(b). A decrease in $|\boldsymbol{c}_{g_1}|$ causes $\wave_1$ to remain within the spatial confines of $\wave_0$ for longer and hence increases the time in which it can extract energy. The red and green shaded regions on Figure \ref{fig: interaction time}(b), associated with $|\boldsymbol{c}_{g_1}|$ and $|\boldsymbol{c}_{g_2}|$, respectively, mark the range of non-dimensional group velocities contained within the secondary beams due to their broadband spectrum. While there is only one wavenumber vector $\boldsymbol{k}_1$ at each $\omega_1$ that can satisfy both dispersion (\ref{eqn: dispersion}) and the triadic resonant condition (\ref{eqn:  phase resonant_condition}), (shown by the loci on Figure \ref{fig: loci_curve}), both $\wave_1$ and $\wave_2$ are beams that are broadly distributed over the wavenumber spectrum due to their finite-width and so there will be exact triads selected from this distribution. A wavenumber distribution was clearly shown experimentally in Figure \ref{fig: wavenumber spectrograms}, where a spectrum of $l_1$ and $l_2$ were observed, changing in both the physical location and duration of the experiment. As $|\boldsymbol{c}_{g_1}|$, defined in (\ref{eqn: cg}), is a function of wavenumber, this spectrum strongly impacts the range of group velocities present in the beam. The strength of the shading in Figure \ref{fig: interaction time}(b) corresponds to the amplitude of the power spectrum for each wavenumber, obtained from Fourier transforming each secondary wave beam profile. The analytically calculated profile assumes a sinusoid with characteristic wavenumber given by the solid branch of the loci in Figure \ref{fig: loci_curve} and width given by the geometry of the triad (assuming a primary beam width $\Lambda_{0}$), enclosed in a Gaussian envelope. Various windowing functions were tested and were not found to significantly alter the range of wavenumbers obtained. 

The result of these two changing factors, $\delta_q$ and $|\boldsymbol{c}_{g_q}|$ (where $q = 1$ or 2), shown in Figure \ref{fig: interaction time}(a) and (b), respectively, are combined to form a residence time $R_q =  \delta_q / |\boldsymbol{c}_{g_q}|$ that characterises how long each secondary wave beam spends within $\wave_0$. The non-dimensional residence time, given as a function of frequency, is shown in Figure \ref{fig: interaction time}(c). As $\omega_1$ increases, $R_1$ also increases. While an increase in $\omega_1$ results in a decrease to $\omega_2$ and therefore a shallower angle for $\wave_2$, as $\wave_2$ and $\wave_0$ are propagating the same direction, the residence time $R_2$ is always greater than $R_1$ and it is never the limiting factor in the interaction. This is shown by the consistently larger values of $R_2$ compared to $R_1$.

Overlaid on Figure \ref{fig: interaction time}(c) are the inverse of the linear growth rates $\sigma$ (given by the form $e^{\sigma t}$), which are marked for all 36 simulations shown in Figure \ref{fig: amp35_outerbranch} by the same $\pair_{\Phi}$ marker style as Figure \ref{fig: loci_curve}, along with many others for different simulations, marked with a hexagon. The linear growth rates are obtained by a linear fit on a logarithmic-linear plot to the initial growth of each simulation. The inverse growth rate can be viewed as a `development time', that characterises how long the secondary beams take to grow. The colour of each mark indicates which behavioural evolution the simulation corresponds to. The magenta is used when no instability arose, the black markers represent those simulations with amplitude modulations and the blue marks indicate the simulations that achieved steady state. The behaviours of the simulations (shown with different style markers and larger size) can be verified from Figure \ref{fig: amp35_outerbranch}.

Figure \ref{fig: interaction time} suggests why the $\Mt$ model is so sensitive to the secondary wave parameters and why three different behavioural evolutions are observed across the triadic configurations tested. For the cases where no instability occurs, marked in magenta on Figure \ref{fig: interaction time}(c), the development time of the secondary beams is greater than the residence time, meaning $\wave_1$ propagates out of $\wave_0$ before sufficient energy transfer can occur. This is case for all the input amplitudes shown in Figure \ref{fig: amp35_outerbranch}(a) and (b) using the triad configurations $\pair_a$ and $\pair_b$, where no growth of the secondary wave beams is observed. In this case, $R_1\sigma < 1 $. 

For the triad configurations and amplitudes that exhibited the quasi-periodic modulations in Figure \ref{fig: amp35_outerbranch}, their development times are marked in black. The majority of these points lie within the range of residence times for $\wave_1$. For these cases, therefore, the development time of the secondary wave beams is comparable with the time taken for $\wave_1$ to propagate across $\wave_0$. The secondary beams are able to grow but have insufficient time to saturate to a stable equilibrium, as wave perturbations will have moved out of the interaction region before this can occur. Moreover, due to the range of group velocities present in the beam, energy will be leaving the primary beam at different times, enhancing the modulations. For these cases, $R_1\sigma \approx 1$. Using this $R_1\sigma$ measure allows us to account for the forcing amplitude of the primary beam (which affects $\sigma$). For example, for the simulations marked with hexagons at low values of $\omega_1$, the forcing amplitude was significantly increased in order to get the secondary beams to grow. 

The final behavioural evolution, observed for the higher amplitude forcing in Figure \ref{fig: amp35_outerbranch}(e) and (f), is when the secondary beams grow and no modulations are observed; rather we see a smooth, rapid decay of the primary beam and the system reaching a steady state. For these cases, where $R_1\sigma > 1$, the development times (marked by the blue triangle and cross) are sufficiently short compared to the range of residence times $R_1$. This means $\wave_1$ is able to extract sufficient energy from the primary beam to reach a steady equilibrium (set by the value of the non-linear interaction term $I_p$) before exiting the underlying $\wave_0$. This reflects how the system would act in the limit of a plane-wave, where the triadic interactions occur infinitely over space and time and the residence time is always greater than the development time.

Through only considering discrete triadic configurations in the $\Mt$ model we have been able to understand the sensitivity of the instability to the secondary wave parameters and how, in the context of a finite-width beam, the spatio-temporal configuration of the triad plays a fundamental role in the evolution of the instability. The $\Mt$ model shows that when the development time has a comparable time-scale to the duration of residence of a wave-packet, the secondary beams are able to grow but are unable to extract sufficient energy to reach a saturated equilibrium state, resulting is continuous amplitude modulations. This phenomenon is also seen experimentally, yet here there is a whole range of perturbations present in the underlying flow. This means that at different locations in physical space, separate triad configurations will be selected, based on the varying structure of $\wave_0$ across the domain and a range of background perturbations. As one triadic interaction decays, another forms, but this time at a different physical location with modified secondary beam parameters. This explains why, experimentally, modulations were observed not only in physical space but also in Fourier space. In the $\Mt$ model simulations presented here, the interaction region of the triad did not move in physical space, as, by construction, $\wave_1$ and $\wave_2$ have fixed frequencies and wavenumbers, causing growth in a specific location.

\section{Conclusions} \label{sec: conclusions}

Novel experimental results have shown that, when TRI arises in a finite-width internal gravity wave beam, the physical regions containing the secondary wave beams modulate over long time-scales without reaching a steady equilibrium. Analysis using Dynamic Mode Decomposition and Fourier methods show that these modulations are present also in the amplitude of the triadic system and in the Fourier space parameters of the secondary beams. 


Through the development and implementation of a two-dimensional weakly non-linear ($\Mt$) model, we have then been able to dissect the experimental set-up by analysing how individual triad configurations effect the evolution of the instability. This model highlights the importance of considering the instability as a function of space, showing how different frequencies of the secondary wave beams alter both their orientation and their group velocity, hence changing their residence time within the underlying $\wave_0$. By comparing the linear growth rate $\sigma$ of the $\Mt$ model simulations with the residence time of the secondary beams $R_q$, we have identified the conditions under which these modulations appear to occur. When the these two time-scales are comparable (i.e. $\sigma R_q \sim 1$), the secondary beams are able to grow, yet the system is unable to reach an equilibrium state as $\wave_1$ and $\wave_2$ are unable to extract sufficient energy from $\wave_0$. When either the amplitude of $\wave_0$ (increasing $\sigma$) or the residence time $R_2$ is sufficiently increased, the system is able to extract sufficient energy to reach equilibrium.


Through the $\Mt$ model, we have been able to isolate the weakly non-linear dynamics of a single triad in a two-dimensional framework and explain the conditions under which certain triad configurations result in amplitude modulations. Yet, in the $\Mt$ model, when these modulations in amplitude occurred, the spatial location of the instability was fixed in one region of the domain. Experimentally, the reasons we also see modulations in the physical location of the triad and in the Fourier space parameters of $\wave_1$ and $\wave_2$ is due to a whole range of perturbations being present in the underlying flow. A range of perturbations means that at different locations in physical space, different triad configurations with similar linear growth rates will be selected. As one triadic interaction decays another forms, but this time at a different physical location with modified secondary beam parameters. This is why the $\Mt$ model did not exhibit movement of the triad in physical space, since, for the results presented in this paper, $\wave_1$ and $\wave_2$ only correspond to discrete parameters. Further work not reported here \citep{Grayson2021}, suggests that when a host of triad configurations are present in the $\Mt$ model, modulations of the secondary beams are also witnessed in both physical and Fourier space.




While there are many other mechanisms that need to be considered in oceanographic data, understanding the evolution of freely evolving finite-width internal wave beams in unbounded domains is fundamental. Here we have elucidated new developments to our knowledge of how the triadic resonance instability mechanism may manifest in scenarios more akin to those found in the ocean as opposed to monochromatic plane waves.

\section{Acknowledgements} \label{sec: acknowledge}

K. M. Grayson acknowledges support from an Natural Environmental Research Council (NERC) studentship (grant no. NE/L002507/1) and from an Engineering and Physical Sciences Research Council (EPSRC) fellowship (grant no. EP/W522600/1).

For the purpose of open access, the authors have applied a Creative Commons Attribution (CC BY) licence to any Author Accepted Manuscript version arising from this submission.

Declaration of Interests. The authors report no conflicts of interest.


\bibliographystyle{jfm}
\bibliography{references}

\end{document}